\documentclass[fleqn,usenatbib]{mnras}

\usepackage[T1]{fontenc}
\usepackage{ae,aecompl}

\usepackage{graphicx}	
\usepackage{amsmath}	
\usepackage{amssymb}	
\usepackage{upgreek}

\newcommand{\dd}{\ensuremath{\mathrm{d}}}
\newcommand{\DD}{\ensuremath{\mathrm{D}}}
\newcommand{\euler}{\ensuremath{\mathrm{e}}}
\newcommand{\iunit}{\ensuremath{\mathrm{i}}}

\newcommand{\vecx}{\ensuremath{{\bmath{x}}}}
\newcommand{\vecr}{\ensuremath{{\bmath{r}}}}
\newcommand{\veck}{\ensuremath{{\bmath{k}}}}

\newcommand{\vecU}{\ensuremath{{\bmath{U}}}}
\newcommand{\vecnab}{\ensuremath{{\bmath{\nabla}}}}

\newcommand{\jfm}{J.~Fluid~Mech.}

\title[Simulations of merging cool-core clusters]{Viscosity, pressure, and support of the gas in simulations of merging cool-core clusters}

\author[W. Schmidt et al.]{
        W.~Schmidt$^{1,2}$\thanks{E-mail:wolfram.schmidt@uni-hamburg.de}, 
        C.~Byrohl$^{2}$,
        J.~F.~Engels$^{2}$,
        C.~Behrens$^{2}$
        J.~C.~Niemeyer$^{2}$\\
$^{1}$Hamburger Sternwarte, Universit\"at Hamburg, Gojenbergsweg 112, D-21029 
Hamburg, Germany\\
$^{2}$Institut f\"ur Astrophysik, Universit\"at G\"ottingen, Friedrich-Hund-Platz 1, D-37077 G\"ottingen, Germany
}

\date{Accepted version}

\pubyear{2017}

\begin{document}
\label{firstpage}
\pagerange{\pageref{firstpage}--\pageref{lastpage}}
\maketitle

\begin{abstract}
Major mergers are considered to be a significant source of turbulence in clusters. We performed a numerical simulation of a major merger event using nested-grid initial conditions, adaptive mesh refinement, radiative cooling of primordial gas, and a homogeneous ultraviolet background. By calculating the microscopic viscosity on the basis of various theoretical assumptions and estimating the Kolmogorov length from the turbulent dissipation rate computed with a subgrid-scale model, we are able to demonstrate that most of the warm-hot intergalactic medium can sustain a fully turbulent state only if the magnetic suppression of the viscosity is considerable. Accepting this as premise, it turns out that ratios of turbulent and thermal quantities change only little in the course of the merger. This confirms the tight correlations between the mean thermal and non-thermal energy content for large samples of clusters in earlier studies, which can be interpreted as second self-similarity on top of the self-similarity for different halo masses. Another long-standing question is how and to which extent turbulence contributes to the support of the gas against gravity. From a global perspective, the ratio of turbulent and thermal pressures is significant for the clusters in our simulation. On the other hand, a local measure is provided by the compression rate, i.e.\ the growth rate of the divergence of the flow. Particularly for the intracluster medium, we find that the dominant contribution against gravity comes from thermal pressure, while compressible turbulence effectively counteracts the support. For this reason it appears to be too simplistic to consider turbulence merely as an effective enhancement of thermal energy.
\end{abstract}




\section{Introduction}

A sizeable fraction of observed galaxy clusters belongs to the class of cool-core clusters, characterised by density and temperature profiles with pronounced central cusps \citep{SandPon06,ChenReip07,JohnPon09,PrattArn10}. Various mechanisms have been suggested to stabilise cool cores or to lift them into the regime on non-cool cores. By now, there is a consensus that feedback from active galactic nuclei (AGN) plays a key role in preventing the overcooling of clusters, bringing properties of simulated clusters closer to observations \citep[e.g.][]{NamNul07,DubDev10,BorKravt11,MarTeys12,YangSut12,VazzaBruegg12,GasBrig14,LiBry15,PlanFab17,HahnMar17}. It was also suggested, however, that major mergers may significantly impact the energy budget of clusters and at least partially contribute to the heating of cool cores \citep[see][]{SkorHall13,HahnMar17}. Observationally, this is indicated by disturbed clusters, in which cool cores tend to be underrepresented and weaker \citep{SandEdge09,PrattArn10}.

In this article, we revisit the problem of major mergers of cool-core clusters from the perspective of dynamical processes in numerical simulations. We consider the limiting case of gas dynamics with radiative cooling and a homogeneous UV background based on the simulations reported in \citet{SchmEngl16}, as summarised in the next section. Since no local feedback is applied, gravitational heating, shock compression, and turbulent dissipation are the major sources of thermal energy. To enhance the spatial resolution, we simulated a particular instance of a major merger with adaptive mesh refinement (AMR) on top of fixed nested grids in a selected region. Rather than aiming at detailed predictions of observed cluster properties, which remains challenging even if sophisticated subgrid recipes for AGN and stellar feedback are applied, we concentrate on fundamental properties of the gas in clusters, particularly the intracluster medium (ICM) and the warm-hot intergalactic medium (WHIM).

First, we address the question of whether the viscosity, which is highly sensitive to the gas temperature, admits a sufficiently broad range of scales for developed turbulence. From the microphysical point of view, the situation is extremely complex. As pointed out by \citet{EganShea16}, the mean free path of ions and electrons can reach tens or even hundreds of kpc, which would render the fluid approximation invalid even in moderately resolved simulations of clusters. However, interactions between particles are mediated between the magnetic field which is known to be present in clusters. Although the magnetic field is dynamically subdominant, it can be considered as the agent that makes the ICM (and, possibly, also the WHIM) behave like a fluid. This is also reflected by the viscosity of a fully ionised plasma, which was calculated in the non-magnetic case by \citet{Bragin58}. Simple estimates show that the non-magnetic viscosity in the ICM would be so sufficiently high to suppress turbulence \citep{ParrCourt12,RoedKraft13}. This is at odds, however, with observational indications of the ICM being turbulent \citep[see reviews by][]{DolBy08,KravtBor12}. The solution is once more an effect introduced by the magnetic field: diffusive transport is vastly reduced in transversal directions \citep{ParrCourt12,KunzBog12,Hopkins17}. However, it is not entirely clear to which extent this reduces the viscosity of the gas in the fluid-dynamical picture and whether an effective isotropic viscous coefficient is appropriate at all. Having performed only hydrodynamical simulations without explicit viscous dissipation, we cannot resolve this issue here. But by postprocessing our data we are able to calculate the viscosity and the associated viscous dissipation scale (also know as Kolmogorov scale) resulting from various assumptions, which bracket the possible outcomes. In contrast to previous estimates, the subgrid-scale model applied in our simulations allows us to compute the viscous dissipation scale directly from the local rate of energy dissipation (which is a closure for the rate of viscous dissipation in the infinitely resolved Navier-Stokes equation; see \citealt{SchmAlm14}). This analysis is presented in Section~\ref{sc:visc}. 

In Section~\ref{sc:press}, we review different definitions of the turbulent pressure. Guided by incompressible isotropic turbulence, a definition based on the power spectrum might be regarded as obvious choice \citep{ZhuFeng10,ZhuFeng11}. However, this introduces several difficulties in the case of cluster turbulence. Most notably, the cumulative power spectrum for any given wave number $k$, which corresponds to the turbulent pressure on the length scale $\ell\sim 1/k$, depends on the spatial location. This is a consequence of the inhomogeneity of the turbulent flow, which manifests itself in the radial dependence of various velocity statistics \citep{RyuKang08,LauKravt09,PaulIap11,VazzaBrun11,ParrCourt12b,Miniati2014,SchmEngl16,VazzaWitt16}. For this reason, we propose a definition on the basis of the spatially varying turbulent velocity dispersion in real space. To compute this quantity, we apply the method of adaptive Kalman filtering introduced in \citet{SchmAlm14,SchmEngl16}. A dimensionless parameter of interest is the ratio of turbulent and thermal pressures or, equivalently, the square of the turbulent Mach number. While this parameter is related to the degree of compressibility of turbulence, it is also commonly considered as a measure for the relative importance of turbulence for the support of the gas against gravity. This might apply to the average effect of turbulence in clusters
\citep{LauKravt09,VazzaBrun11,ParrCourt12b,MiniBer15,SchmEngl16,VazzaWitt16}. \citet{BifBor16} evaluated shell-averaged accelerations and their deviation from hydrostatic equilibrium under the assumption of spherical symmetry. Although this is questionable for the strongly perturbed clusters during a major merger, their results indicate significant contributions from random gas motions. A different approach was put forward by \citet{ZhuFeng10}, who considered the growth rate of the divergence of the peculiar velocity as a purely local measure for the relative importance of thermal pressure, turbulence, and gravity. This is possible because the gradient of the potential occurring in the momentum equation is converted into a source term of the Poisson equation, which is simply the local density fluctuation times a constant coefficient. The analysis of \citet{ZhuFeng10,ZhuFeng11,IapSchm11} suggests that vorticity enhances the support of the gas. However, they did not fully account for the contributions from shocks, which obviously have a strong impact on the time evolution of the velocity divergence. In Section~\ref{sc:support}, we take up these studies to develop a more complete picture of the role of turbulence in supporting the gas in clusters, followed by our conclusions in Section~\ref{sc:concl}.

\section{Numerical methods and models}

In the following, we give a brief overview of the numerics and models applied in our simulations, which were performed with the cosmological AMR code \textsc{Nyx}. For details, we point the reader to the method and application papers \citet{AlmBell13,SchmAlm14,LukStark15,SchmEngl16}.

\textsc{Nyx} implements a dimensionally unsplit higher-order method to solve the equations of gas dynamics \citep{miller-colella}, a kick-drift-kick method based on \citet{miniati-colella} for $N$-body dynamics, and the \textsc{BoxLib} multigrid algorithm in combination with cloud-in-cell interpolation for particles to solve the Poisson equation for the gravitational potential \citep[see][]{AlmBell13}. A time-varying homogeneous ultraviolet background following \citet{HaardtMad12} and radiative cooling for primordial gas were added by \citet{LukStark15}. Moreover, \citet{SchmAlm14} implemented a subgrid-scale (SGS) model for the numerically unresolved fraction of turbulent kinetic energy. The complete system of equations for the state variables in co-moving coordinates can be found in Section 3 of \citet{SchmEngl16}.

Apart from adding the turbulent stresses associated with energy transfer across the grid scale to the equations of gas dynamics, the SGS model enables us to improve the correction of energy variables when grids are refined or de-refined \citep[see][]{SchmAlm14}. This minimises artificial manipulations of the internal energy, which are necessary to ensure global energy conservation on unstructured grids. Moreover, the turbulent diffusion of internal energy on subgrid scales is incorporated. Since turbulence in the cosmic web is neither homogeneous and nor stationary, \citet{SchmAlm14} introduced a shear-improved SGS model to determine the growth rate of SGS turbulence energy from the shear of numerically resolved velocity fluctuations relative to a temporally smoothed mean flow. The smoothing algorithm is based on the adaptive Kalman filtering technique \citep{CahuBou10}, which encompasses an estimator for smoothing scales depending on the local flow evolution (a detailed description is given in Appendix A.3 of \citealt{SchmAlm14}). An additional advantage of the shear-improved model is the utilisation of the resolved turbulent velocity fluctuations for estimating the total turbulent pressure (see Section~\ref{sc:press}).

\begin{figure*}
\includegraphics[width=\linewidth]{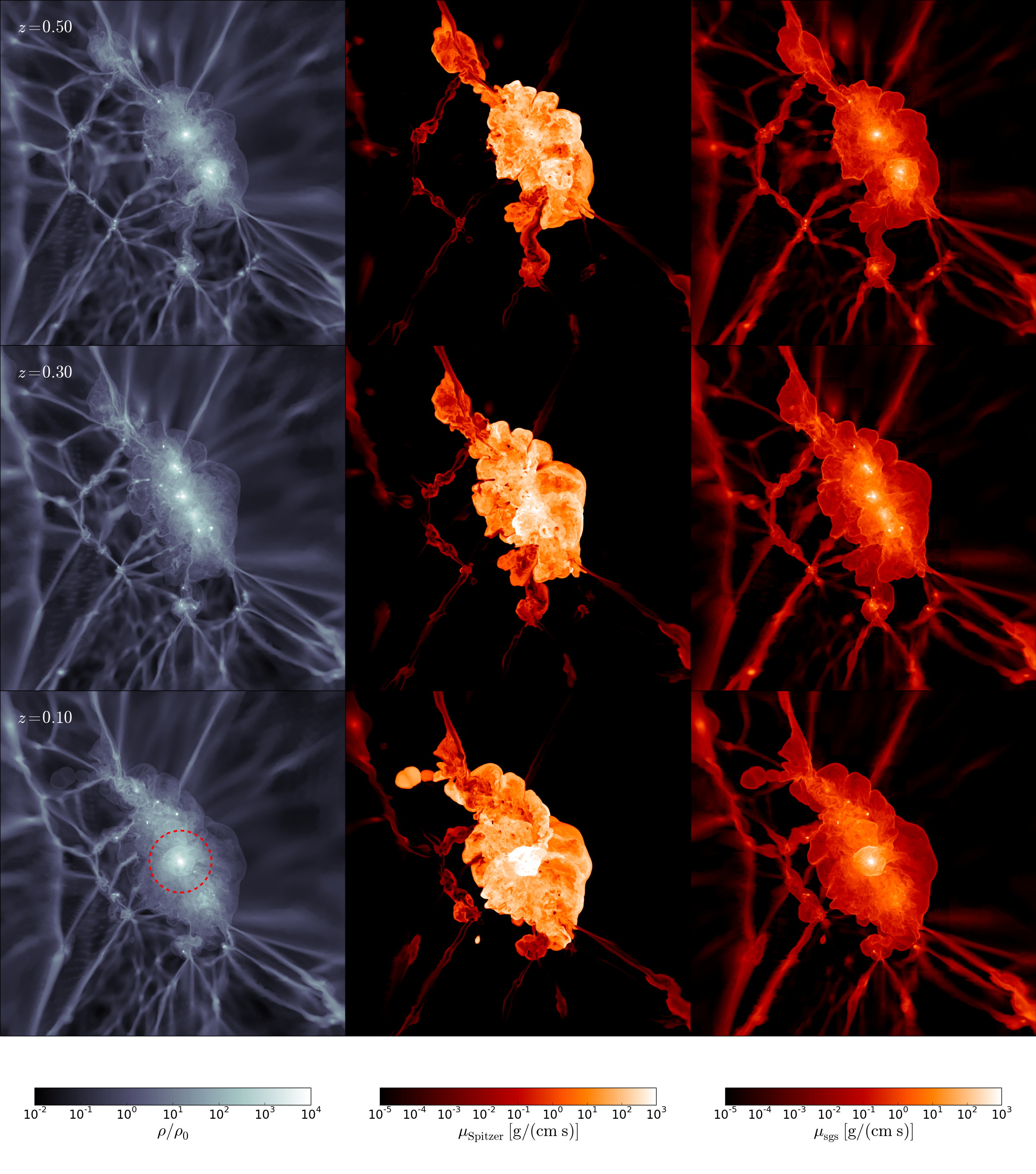}
\caption{Slices of the gas overdensity $\rho/\rho_0$ (left), the Spitzer viscosity $\mu_{\rm Spitzer}$ (middle), and the subgrid-scale viscosity $\mu_{\rm sgs}$ (right) at three different redshifts. Shown is only the nested-grid region, whose size is one quarter of the total box size, i.e.\ 37.9\;{Mpc}. The region is centred at the final density peak at $z=0$. The size of the merged halo determined by the HOP finder is indicated by the red dashed circle. The radial profiles shown in the following figures were computed from the time-dependent density peak outwards to a maximal radius of $R_{\rm max}=15\;\mathrm{Mpc}$, corresponding to a sphere that fits within the nested-grid region. The slices are oriented perpendicular to the slices shown in Fig.~1 of \citet{SchmEngl16}.}
\label{fig:slices_evol}
\end{figure*}

\citet{SchmEngl16} used the cosmological initial condition generator \textsc{Music} \citep{HahnAbel11} to compute particle displacements for a cosmological box of $152\;\mathrm{Mpc}$ co-moving size with cosmological parameters from the \citet{Planck14}. The box was then evolved with \textsc{Nyx} from redshift $z=99$ to $0$. Global refinement by overdensity and vorticity made the simulations extremely resource-intensive, limiting their resolution to $78\;\mathrm{kpc}$. To improve on that and to increase the mass resolution at least for a selected cluster, we centred the box at the density peak of the most massive halo (halo 1 in Table 4 in \citealt{SchmEngl16}; the halo mass is $M_{\rm halo}=6.68\times 10^{14}\,M_\odot$, its maximal radial extent $R_{\rm halo}=3.24\;\mathrm{Mpc}$, and $R_{500}=1.04\;\mathrm{Mpc}$). Further analysis revealed that this cluster is the result of a major merger at low redshift. We generated nested-grid initial conditions for $512^3$ particles/cells at the root-grid level and two nested-grid levels extending over the central quarter of the box in each spatial direction (plus extensions enforced by buffer zones between different refinement levels). The second nested-grid level has about the same number of particles and resolution elements as the root-grid level, but four times higher spatial resolution and 64 times smaller particle mass, resulting in a substantially improved representation of subhalos. On top of that two levels of AMR were applied within the nested-grid region, using the same refinement criteria as in the global simulation. Thus, the maximal spatial resolution is $18.5\;\mathrm{kpc}$. In terms of memory requirement (more than $10^9$ grid cells at the highest refinement level), the simulation is comparable to run $\text{B}_2$ of \citet{SchmEngl16}. Owing to the larger number of hydrodynamical time steps imposed by the CFL criterion and the extensive subcycling for the treatment of cooling, particularly in galaxy-mass halos, the required computation time was much longer. Nevertheless, we were able to advance the nested-grid/AMR simulation all the way to redshift zero and to carry out the analysis presented in the following sections. An impression of the cluster residing in the central halo and the preceding merger is given by the density slices across the nested-grid region in Fig.~\ref{fig:slices_evol}.

\section{Viscosity}
\label{sc:visc}

In principle, there are three different viscosities that come into play in numerical simulations of turbulent flows.
Firstly, the microscopic viscosity is a material property of the fluid and significant on length scales comparable to or smaller than the Kolmogorov scale $\ell_{\rm K}$. Its effect is the physical dissipation of kinetic energy into heat. Secondly, the turbulent or, more precisely, subgrid-scale viscosity is a property of the physical flow below the chosen grid scale $\Delta$ of the simulation in the limit of high Reynolds numbers. The SGS viscosity causes a transfer of kinetic energy from numerically resolved to unresolved scales. Thirdly, the numerical viscosity is purely a property of the numerical scheme that is applied to approximate the flow numerically. Its main purpose is the stabilisation of the numerical solver. A side effect of numerical viscosity is the artificial dissipation of kinetic energy at length scales comparable to the grid resolution.

Depending on the ratio of the viscosities, two fundamentally different simulation regimes are encountered (for a detailed discussion see \citealt{Schmidt15}):
\begin{itemize}
\item Fully resolved or direct numerical simulation: The Kolmogorov scale is within the range of numerically accessible scales (i.e.\ $\ell_{\rm K}\gtrsim\Delta$). In this case, the Navier-Stokes equations with an explicit viscous stress tensor have to be solved. In astrophysics, this is only rarely possible. Examples, particularly in the context of the ICM, are \citet{ParrCourt12,KunzBog12,RoedKraft13,RoedKraft15}. A code with advanced treatment of both isotropic and anisotropic viscosity was recently put forward by \citet{Hopkins17}.
\item Under-resolved or large eddy simulation (LES): If the Kolmogorov scale is much smaller than the grid scale ($\ell_{\rm K}\ll\Delta$), energy dissipation is shifted from the Kolmogorov to the grid scale by numerical viscosity (implicit LES) or treated via a subgrid-scale model. Implicit LES (ILES for short) formally solve the Euler equations. However, the numerical viscosity introduced by finite volume schemes, which are commonly used for astrophysical fluid dynamics, mimics the effect of viscous dissipation. As a result, the numerically computed flow effectively behaves like Navier-Stokes turbulence in the inertial subrange, for which the nature and value of the viscosity is unimportant. This approach is used in the majority of cosmological simulations. The simulations considered here are LES with an explicit SGS viscosity. Apart from the explicit coupling of numerically resolved and unresolved scales, this allows us to compute the local dissipation rate (i.e., the thermal energy produced by turbulent dissipation per unit time; see \citealt{SchmFeder11,SchmAlm14}) as 
\begin{equation}
	\label{eq:diss}
	\varepsilon = \frac{C_\varepsilon K^{3/2}}{\Delta},
\end{equation}
where $K$ is the subgrid-scale energy per unit mass and $C_\varepsilon$ a dimensionless coefficient, which is set equal to $2^{3/2}\times 0.56$.
\end{itemize}
 
\begin{figure*}
	\includegraphics[width=\linewidth]{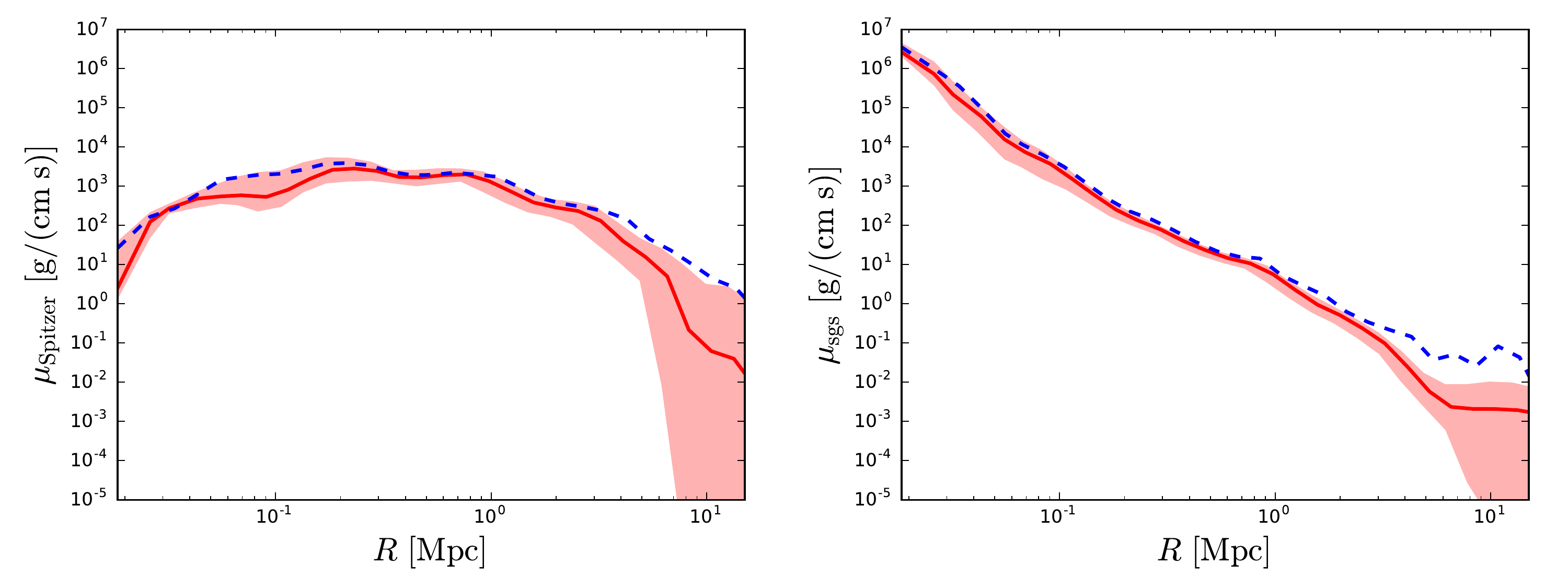}
    \caption{Comparison of the dynamic Spitzer viscosity (left) and subgrid-scale viscosity (right). Shown are volume-weighted radial profiles (blue dashed lines), medians (red lines), and the interquartile range (red shaded regions). }
    \label{fig:visc}
\end{figure*}

\begin{figure*}
	\includegraphics[width=\linewidth]{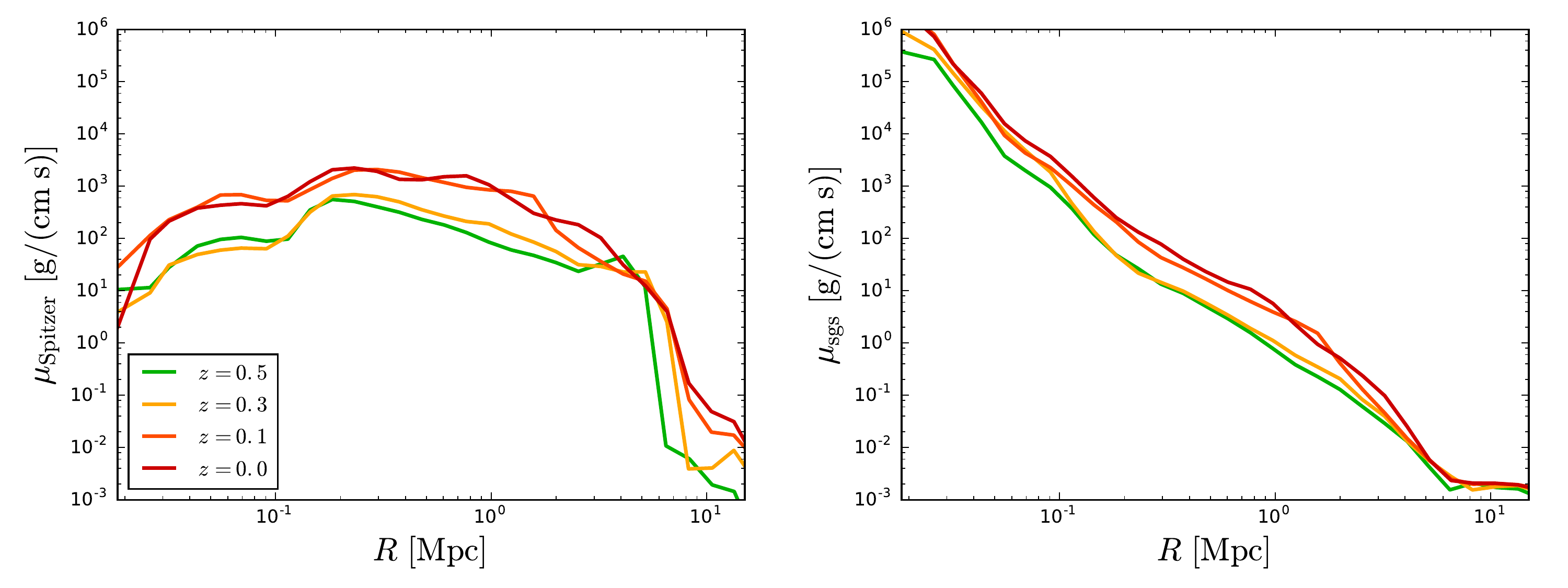}
    \caption{Medians of the Spitzer viscosity (left) and subgrid-scale viscosity (right) at different redshifts. }
    \label{fig:visc_evol}
\end{figure*}

The dynamic viscosity of a fully ionised gas in the absence of magnetic fields \citep[see][]{Spitzer62},\footnote{
	Actually, this result was originally obtained by \citet{Bragin58}, but the term Spitzer viscosity is quite common in the literature.}
\begin{equation}
	\label{eq:visc_spitzer} 
	\mu_{\rm Spitzer} = 
	2.21\times10^{-15}\,\frac{(T\,[\mathrm{K}])^{5/2}A_{\rm i}^{1/2}}{Z^4\ln\Lambda}\;\mathrm{g\,cm^{-1}\,s^{-1}}\,,
\end{equation}
is mainly a function of the temperature $T$. The parameters $A_{\rm i}$ and $Z$ are the atomic weight and charge, respectively, of the ions. The Coulomb integral $\ln{\Lambda}$ depends logarithmically on temperature and density, and is commonly approximated by a constant value around $40$ for clusters \citep[see][]{GasChur13,SmithShea13}. Slices of the resulting viscosity are shown in middle panels of Fig.~\ref{fig:slices_evol} for three stages of the simulated merger. In Fig.~\ref{fig:visc}, radial profiles of volume-averages (blue), medians (red), and interquartiles (red shaded) are plotted for the cluster produced by the merger at $z=0$. The typical value of the Spitzer viscosity in the WHIM is $\mu_{\rm Spitzer}\sim 10\;\mathrm{g\,cm^{-1}\,s^{-1}}$, with a gradual increase toward the core. Around the accretion shock ($R\approx 5\;\mathrm{Mpc}$), the viscosity drops rapidly. It turns out that the median profile reflects this drop much better than the volume-averaged (or mass-averaged) profile because averaging a quantity that varies over several orders of magnitude is biased toward high values. Beyond the outer shocks, the values calculated with equation~(\ref{eq:visc_spitzer}) become physically meaningless because the unshocked gas is neutral or only partially ionised. For the ICM, viscosities above $100\;\mathrm{g\,cm^{-1}\,s^{-1}}$ would be representative if it were not for the depression within $100\;\mathrm{kpc}$ from the centre. As discussed in \citet{SchmEngl16}, this is a consequence of the over-pronounced cool cores in our simulations. Feedback, particularly from AGNs, would rise the core temperature and, thus, the viscosity. This is important to bear in mind for the following discussion. In any case, the viscosity following from equation~(\ref{eq:visc_spitzer}) is large enough to significantly affect or even suppress dynamical instabilities in both the ICM and the WHIM. This was also demonstrated by \citet{RoedKraft13,RoedKraft15,ZuHone15} in the context of Kelvin-Helmholtz instabilities and gas sloshing induced by minor mergers.

\begin{figure*}
	\includegraphics[width=\linewidth]{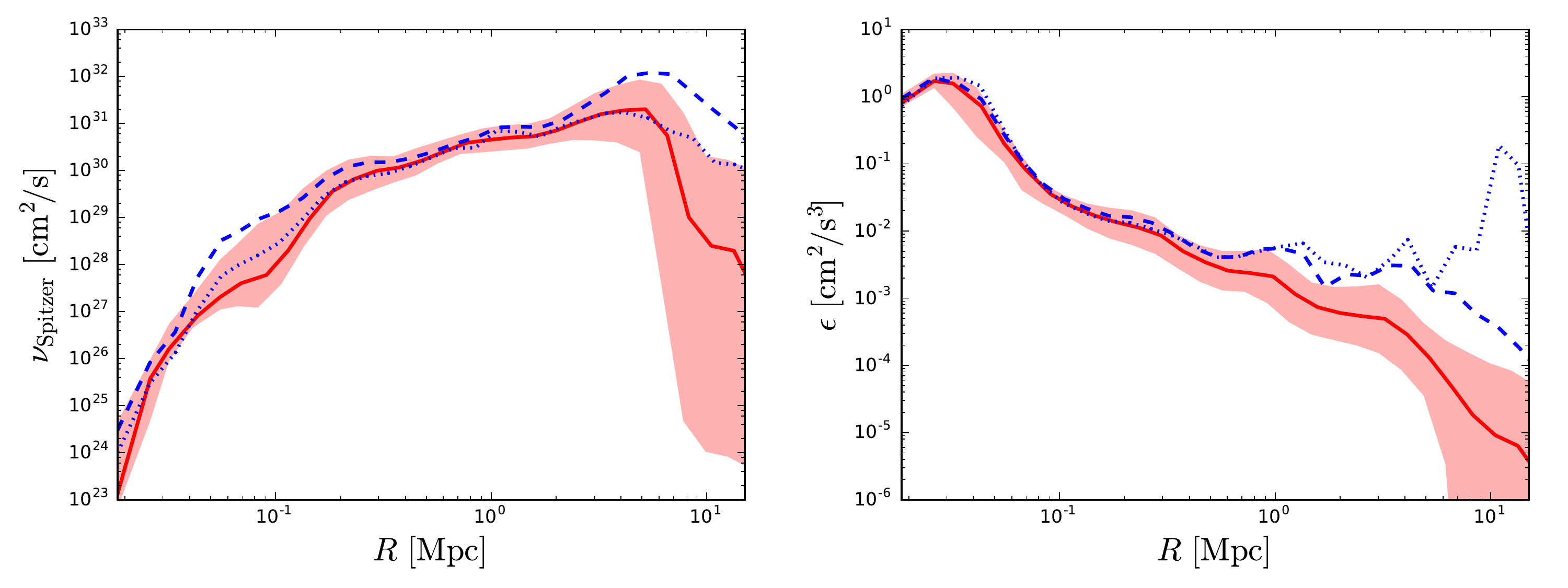}
    \caption{Radial profiles (mass-weighted: blue dotted, volume-weighted: blue dashed, median: red solid, interquartile: red shaded) of the kinematic Spitzer viscosity (left) and turbulent dissipation rate (right).}
    \label{fig:visc_kin}
\end{figure*}

Assuming the Spitzer viscosity as worst possible case, it is an important question whether the underlying assumption of the LES approach is valid. In terms of viscosities, this assumption is equivalent to $\mu_{\rm Spitzer}\ll\mu_{\rm sgs}$, where the subgrid-scale viscosity is given by
\begin{equation}
	\label{eq:visc_sgs} 
	\mu_{\rm sgs} = \rho C_{\nu}\Delta\sqrt{K} 
\end{equation}
with the scale-free coefficient $C_{\nu} \approx 0.1$ \citep[see][]{Schmidt15}. However, if the profiles of $\mu_{\rm Spitzer}$ and $\mu_{\rm sgs}$ in Fig.~\ref{fig:visc} are compared, this is clearly not the case. Particularly in the WHIM, at radii of a few Mpc, the Spitzer viscosity turns out to be much higher. This is also quite obvious from the slices shown in Fig.~\ref{fig:slices_evol}. The subgrid-scale viscosity exceeds the Spitzer viscosity only in the core region, but this is mainly due to the density factor in equation~(\ref{eq:visc_sgs}) and the low core temperature. It is not clear how feedback changes the picture, as it would both heat up the ICM and stir the gas, resulting in enhanced turbulence. In this regard, the major merger around $z=0.2$ has a similar impact: Fig.~\ref{fig:visc_evol} shows that both the Spitzer and the SGS viscosities are systematically lower prior to the merger. This trend reflects the heating of the gas by merger shocks in and an enhanced production of turbulence.

Equations~(\ref{eq:visc_spitzer}) and (\ref{eq:visc_sgs}) also enable us to estimate the importance of thermal conduction vs.\ the turbulent diffusion of thermal energy, which was introduced in \citet{SchmEngl16}. 
On the one hand, \citet{ParrCourt12} and \citet{KunzBog12} show that the Prandtl number for the microscopic diffusivities,
\begin{equation}
	\mathrm{Pr}=\frac{\nu}{\kappa}
\end{equation}
is about $0.02$, i.e. the thermal conductivity $\kappa$ is two orders of magnitude larger than the kinematic viscosity $\nu=\mu/\rho$. 
The transport of internal energy by numerically unresolved eddies is modelled by a gradient-diffusion closure with the diffusivity
\begin{equation}
	\kappa_{\rm sgs} = C_{\kappa}\Delta\sqrt{K},
\end{equation}
where $C_{\kappa}\approx 0.4$ \citep{SchmNie06}, corresponding to a kinetic Prandtl number of about $0.25$. 
The Prandtl numbers imply $(\kappa_{\rm sgs}/\kappa)\sim 10(\mu_{\rm sgs}/\mu)$. 
This estimate suggests that the impact of conduction relative to turbulent diffusion is even larger than in the case of the viscosities. 
In other words, if the viscosity were given by the Spitzer viscosity, it would immediately follow from the profiles shown in Fig.~\ref{fig:visc_evol} that the transport of heat on scales comparable to the grid resolution would be conduction-dominated in most of the cluster. 

\begin{figure}
	\includegraphics[width=\linewidth]{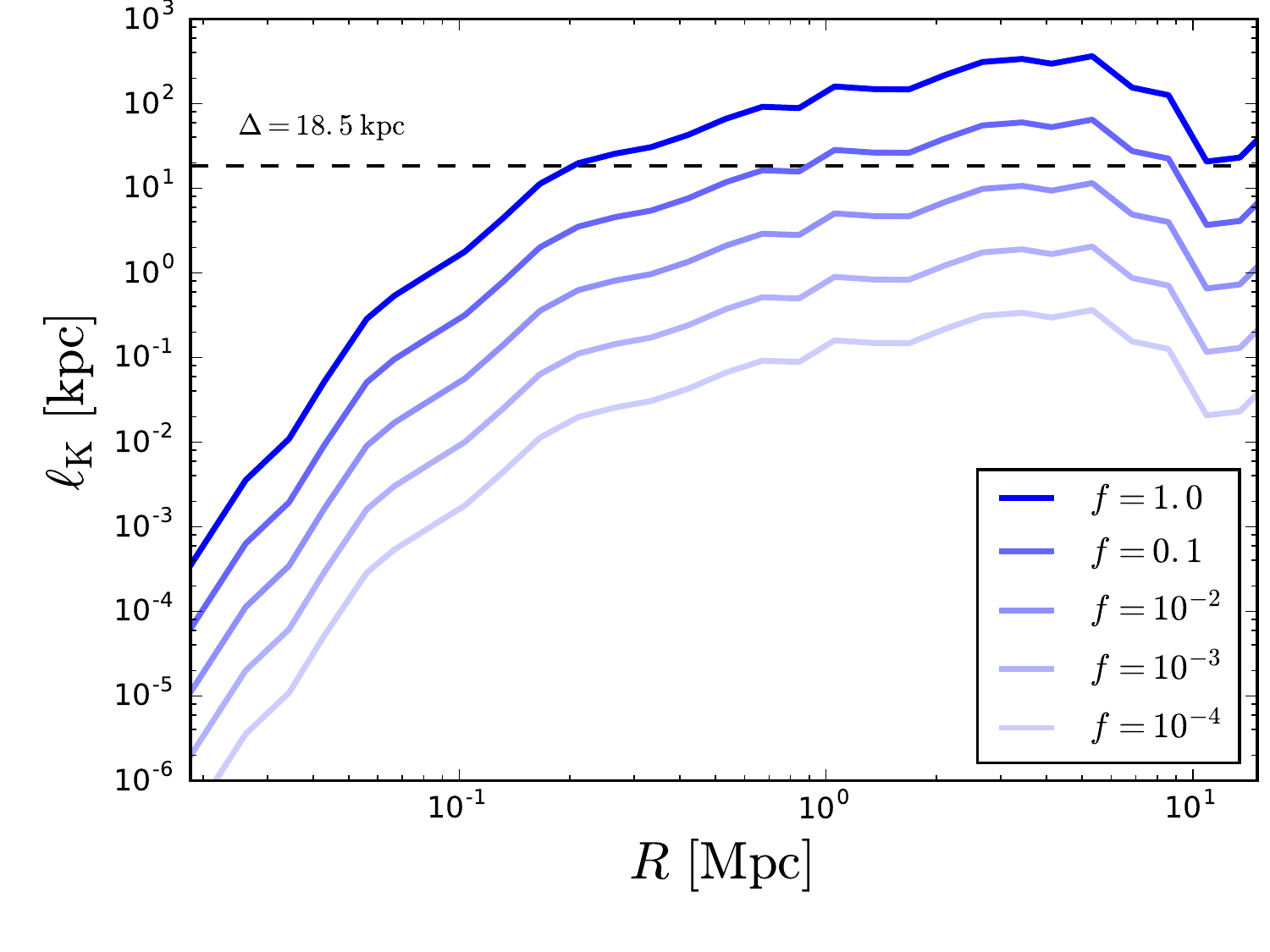}
    \caption{Profiles of the Kolmogorov length scale computed from mass-weighted shell averages of the effective kinematic viscosity, which is assumed to be given by some fraction $f$ of the Spitzer viscosity, and the dissipation rate. The grid scale at the highest refinement level is indicated by the horizontal dashed line.}
    \label{fig:visc_scale}
\end{figure}

\citet{ParrCourt12} and \citet {RoedKraft13} estimate the Reynolds number of the ICM in terms of some characteristic scales of turbulence and the kinematic Spitzer viscosity. Although the turbulent velocity dispersion computed in our simulations yields the required velocity scale (see Section~\ref{sc:press}), the integral length scale at which turbulence is injected is highly uncertain (estimates reach from less than $100\;\mathrm{kpc}$ to about a Mpc; see \citealt{RyuKang08,ParrCourt12,RoedKraft13,GasChur14,SchmAlm14,VazzaBruegg14,Miniati2015}). Owing to the strong density and temperature dependence of the viscosity, very different Reynolds numbers are obtained depending on local conditions. Consequently, there is no typical Reynolds number characterising the turbulent state of the ICM (and even less so the WHIM). Here, we take a different approach and consider the local Kolmogorov length defined by\footnote{One might object that equation~(\ref{eq:kolmog_length}) applies only to incompressible turbulence. Apart from the strong outer shocks induced by accretion, however, turbulence in most of the gas in clusters is typically not supersonic, as \citet{SchmEngl16} and many others have shown. For this reason, the Kolmogorov length can be considered as a reasonable approximation to the length scale of microscopic dissipation for our order-of-magnitude estimation.} 
\begin{equation}
	\label{eq:kolmog_length}
	\ell_{\rm K} = \left(\frac{\nu^3}{\epsilon}\right)^{1/4},
\end{equation}
where $\epsilon$ the dissipation rate defined by equation~(\ref{eq:diss}). Both quantities are well defined field variables in our simulations. The only complication arises from the fact that equation~(\ref{eq:kolmog_length}) applies to an ensemble average. Since $\nu$ and 
$\epsilon$ vary substantially with radial distance from the centre of the cluster (profiles are shown in Fig.~\ref{fig:visc_kin}), we face a similar problem as in the case of the Reynolds number (without the additional difficulty of estimating the driving scale). A sensible procedure to compute profiles of $\ell_{\rm K}$ is to average $\nu$ and $\epsilon$ over radial bins and to substitute the averages into equation (\ref{eq:kolmog_length}). By setting the viscosity to different fractions of the Spitzer viscosity, i.e.~$\nu=\nu_{\rm eff}=f\mu_{\rm Spitzer}/\rho$, the profiles plotted in Fig.~\ref{fig:visc_scale} are obtained. The fraction $f$ is interpreted as magnetic suppression factor \citep[for a more detailed discussion, see][]{GasChur13,SmithShea13}. The effective viscosity $\nu_{\rm eff}$ corresponds to an isotropic viscosity on macroscopic scales, resulting from the anisotropic reduction of the mean free path due to magnetic fields. Our results suggest that a suppression factor $f\sim 10^{-3}$ would be sufficient to satisfy the basic assumption of LES, $\ell_{\rm K}/\Delta\ll 1$, even in the WHIM. This value is within the range of estimates following from theoretical considerations and observational constraints \citep{GasChur13}. An upper limit of $f\approx 0.05$ was deduced by \citet{SuKraft17} from Chandra observations of a minor merger event. This value would correspond to a Kolmogorov scale of the order of the grid resolution in most of the WHIM.

\begin{figure}
	\includegraphics[width=\linewidth]{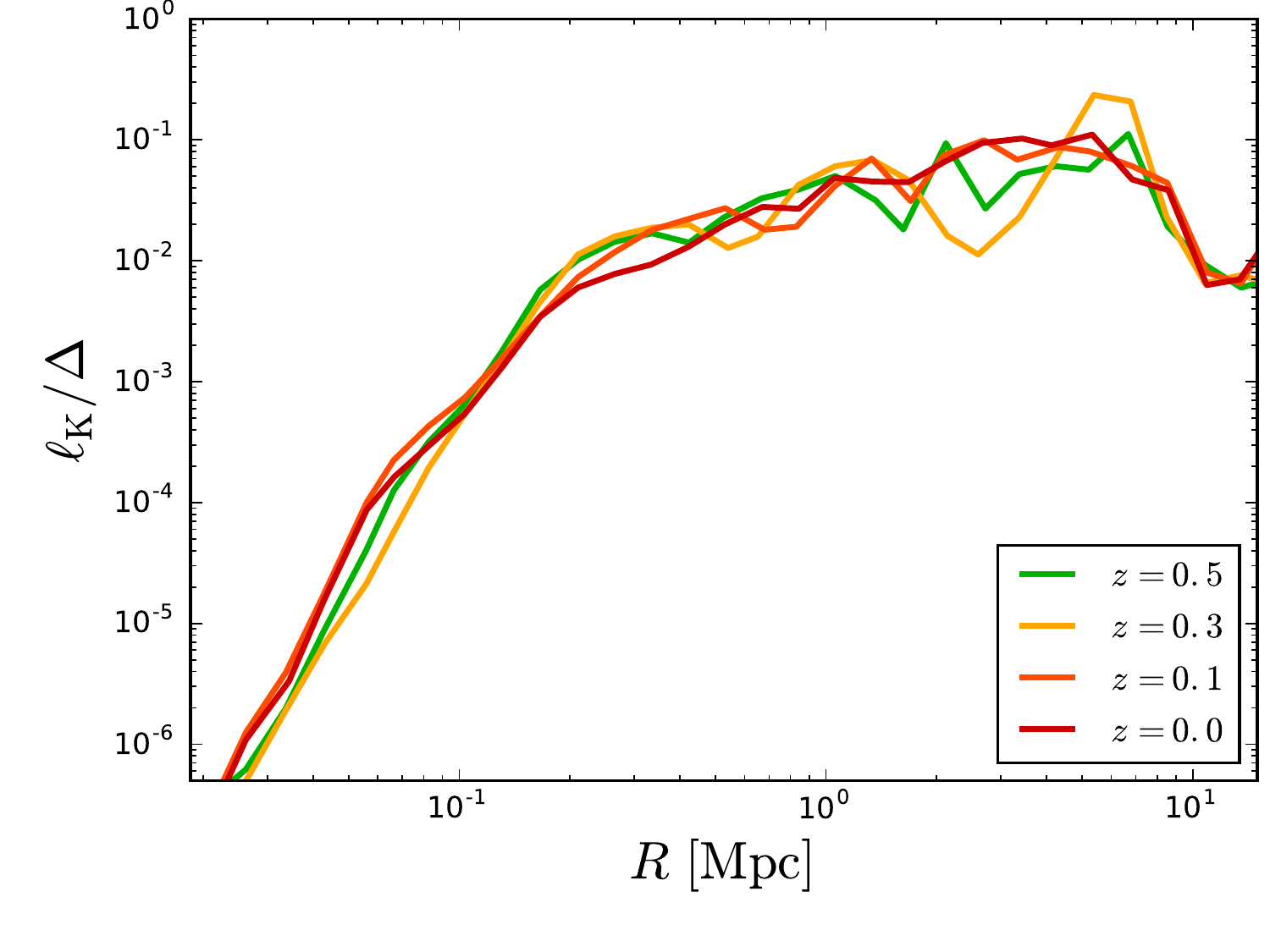}
    \caption{Evolution of the Kolmogorov length relative to the grid scale for $f=10^{-3}$.}
    \label{fig:visc_scale_evol}
\end{figure}

Figure~\ref{fig:visc_scale_evol} shows plots of $\nu_{\rm eff}$ for $f=10^{-3}$ at different redshifts. It turns out that the Kolmogorov scale is largely insensitive to changes in the thermal and turbulent state of the gas, at least in the  range of redshifts considered here. This comes about because the higher viscosity after the merger is roughly compensated by the higher turbulent energy dissipation rate.

\begin{figure*}
	\includegraphics[width=\linewidth]{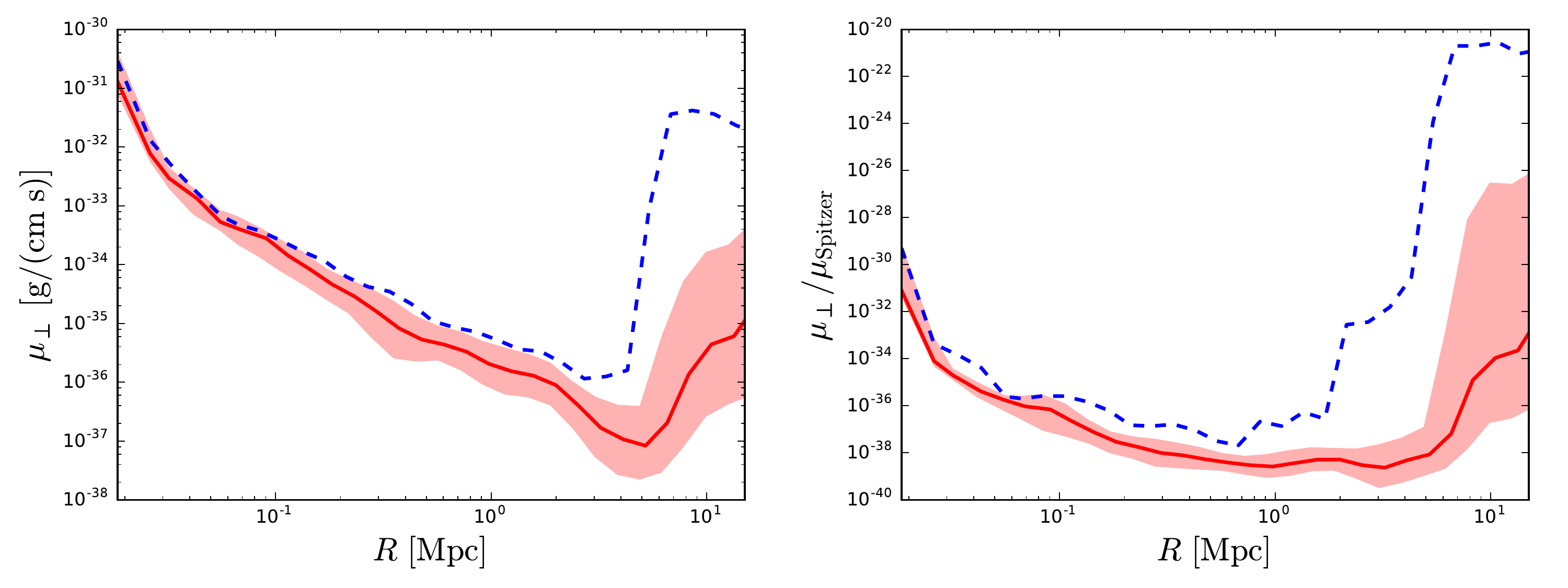}
    \caption{Radial profiles (volume-weighted: blue dashed, median: red solid, interquartile: red shaded) of the transversal viscosity (left) and the ratio of the transversal to the Spitzer viscosity (right) assuming the saturated magnetic field given by equation~(\ref{eq:magn_field_sat}) at redshift $z=0$.}
    \label{fig:visc_magn}
\end{figure*}

For the special case of a flow in transverse direction to the magnetic field, \citet{Simon55} found that the viscosity for shear perpendicular to the plane spanned by the velocity and the magnetic field is given by
\begin{equation}
	\label{eq:visc_trans} 
	\mu_{\perp} = 
	2.68\times10^{-26}\,\frac{A_{\rm i}^{3/2}Z^2 (n_{\rm i}\,[\mathrm{g\,cm^{-3}}])^2\ln\Lambda}{(T\,[\mathrm{K}])^{1/2}(B\,[\mathrm{G}])^2}\;\mathrm{g\,cm^{-1}\,s^{-1}}\,,
\end{equation}
where $n_{\rm i}$ is the ion number density and $B$ the magnetic field strength. This result applies only in the limit $\omega_{\rm ci}t_{\rm ci}\gg 1$, where $\omega_{\rm ci}$ is the cyclotron frequency of ions and $t_{\rm ci}$ their self-collision time, i.e.\ for sufficiently strong magnetic fields. If the velocity component parallel to the field direction induces shear along the field lines, equation~(\ref{eq:visc_spitzer}) applies \citep{Bragin65}. The ratio of the transversal and parallel viscosity coefficient is thus given by
\begin{equation}
	\label{eq:visc_aniso} 
	\frac{\mu_{\perp}}{\mu_{\rm Spitzer}} = \frac{1}{(\omega_{\rm ci}t_{\rm ci})^2} \propto \frac{n_{\rm i}^2}{T^3 B^2}\ll 1.
\end{equation}  
Consequently, the viscosity of a fully ionised gas is strongly anisotropic in the presence of a (strong) magnetic field and requires a tensor representation \citep[see][]{KunzBog12,Hopkins17}. Although the above formulas pertain to special cases, we can nevertheless estimate the impact of magnetic fields in clusters. Since MHD is not applied in our simulations, we approximate the field strength by assuming a certain fraction of energy equipartition in the saturated regime \citep[see][]{SchmAlm14}:\footnote{
	Even if MHD is employed in cosmological simulations, the resulting magnetic field is not necessarily more reliable because the amplification due to the turbulent dynamo tends to be significantly underestimated in under-resolved numerical simulations \citep{VazzaBruegg14,Cho2014,MiniBer15,EganShea16}.}
\begin{equation}
	\label{eq:magn_field_sat} 
	B_{\rm sat}\simeq 2.24\rho^{1/2}\sigma_{\rm turb}.
\end{equation}
in Gaussian units. We find $B_{\rm sat}\gtrsim 1\;\mathrm{\mu G}$ in the ICM (with a field strength well above $10\;\mathrm{\mu G}$ in the core) and a few $100\;\mathrm{nG}$ in the WHIM at $z=0$. By substituting the above expression into equations~(\ref{eq:visc_trans}) and~(\ref{eq:visc_aniso}), the profiles shown in Fig.~\ref{fig:visc_magn} are obtained. Indeed, a tremendous reduction of the transversal viscosity compared to the Spitzer viscosity follows if a saturated magnetic field is assumed. However, the ratio $\mu_{\perp}/\mu_{\rm Spitzer}$ arises from a coherent magnetic field on microscopic scales and must not be confused with the magnetic field suppression factor $f$ for the effective isotropic viscosity, which accounts for the effect of the strongly tangled field in a turbulent plasma.

\section{Pressure}
\label{sc:press}

In the following, we focus on the pressure associated with macroscopic random motions of a fluid, which is called turbulent pressure, and its relation to the thermal pressure induced by microscopic motions of particles. \citet{ZhuFeng10} compute the turbulent pressure associated with velocity fluctuations below a given length scale $l$, corresponding to the wavenumber $k=2\pi/l$, from the energy spectrum function $E(k)$:
\begin{equation}
	\label{eq:press_spect}
	P(k) = \int_k^{\infty}E(k^\prime)\,\dd k^\prime,
\end{equation}
This expression is also known as cumulative energy spectrum. For viscous fluids, the upper limit of the integral is given by $k_{\rm max}=\pi/\ell_{\rm K}$, where $\ell_{\rm K}$ is the Kolomogorov length. However, the above method of calculating the turbulent pressure entails several difficulties. To begin with, there is no unambiguous definition of $E(k)$ for compressible turbulence \citep[see, for example,][]{KritNor07,FederRom10,PietCam10,Aluie11,GalBan11,WagFalk12}. Moreover, the energy spectrum depends on position and time if turbulence is not homogeneous and stationary.For incompressible turbulence, this follows from the general definition of the energy spectrum function as surface integral of the energy spectrum tensor over a sphere with radius $k$ in Fourier space:
\begin{equation}
	  E(\vecx,k,t) = 
	  \frac{1}{2}\oint_{|\veck^\prime|=k}\Phi_{ii}(\vecx,\veck^\prime,t)\,k^{\prime 2}\dd\Omega_{\veck^\prime},
\end{equation}
The energy spectrum tensor $\Phi_{ij}(\vecx,\veck,t)$ in turn is the Fourier transform of the velocity correlation tensor:
\begin{equation}
	  \Phi_{ij}(\vecx,\veck,t) = 
	  \frac{1}{(2\pi)^3}\int \langle U_i^\prime(\vecx)U_j^\prime(\vecx+\vecr)\rangle\euler^{-\iunit\vecr\cdot\veck}\,\dd^3 r
\end{equation}
Brackets denote the ensemble average. The dependence on $\vecx$ and $t$ averages out only in the case of statistically stationary and homogeneous turbulence, for which $E(\vecx,k,t)=E(k)$ is purely a function of the wavenumber.
\citet{ZhuFeng10} claim that turbulence is homogeneous and isotropic in their cosmological simulation. But their reasoning is circular because they use diagnostics based on the assumptions of incompressibility, homogeneity, and isotropy. In their more recent analysis, \citet{ZhuFeng15} find that the vorticity magnitude increases from the void over filaments and sheets to clusters. In \citet{SchmEngl16}, it is demonstrated that turbulence in the cosmic web is highly inhomogeneous, with systematic differences between the ICM and the WHIM, much weaker turbulence in filaments, and the gas in the void being not only dilute, but also largely devoid of turbulence. Naturally, there is strong anisotropy across accretion shocks.

Nevertheless, we attempt to relate our notion of turbulent pressure, which is based on the local magnitude of turbulent velocity fluctuations in physical space, to $P(k)$ defined by equation~(\ref{eq:press_spect}). First we observe  that $P(k)$ asymptotically reaches a value independent of $k$ for $k \ll k_0$, where $k_0=2\pi/L$ is the
wave number of the largest turbulent eddies, whose size is of the order of the integral length scale (also called driving scale) $L$. This limit roughly corresponds to the total turbulent pressure given by the turbulent velocity dispersion $\sigma_{\rm turb}$:
\begin{equation}
	\lim_{k\rightarrow 0}P(k) \sim 
	P_{\rm turb} := \frac{1}{3}\rho\sigma_{\rm turb}^2 = \frac{1}{3}\rho\left(U^{\prime\,2} + 2 K\right)
	\label{eq:press_turb}
\end{equation}
The first term in the expression substituted for $\sigma_{\rm turb}^2$ on the right-hand side is the contribution of numerically resolved turbulent velocity fluctuations $U^\prime$ (in our simulation determined by Kalman filtering; see \citealt{SchmAlm14}). The second term results from subgrid-scale turbulence. While the above expression is a constant in the case of stationary homogeneous turbulence, $P_{\rm turb}$ varies in space and time for the non-stationary inhomogeneous turbulence in the cosmic web. 

\begin{figure*}
	\includegraphics[width=\linewidth]{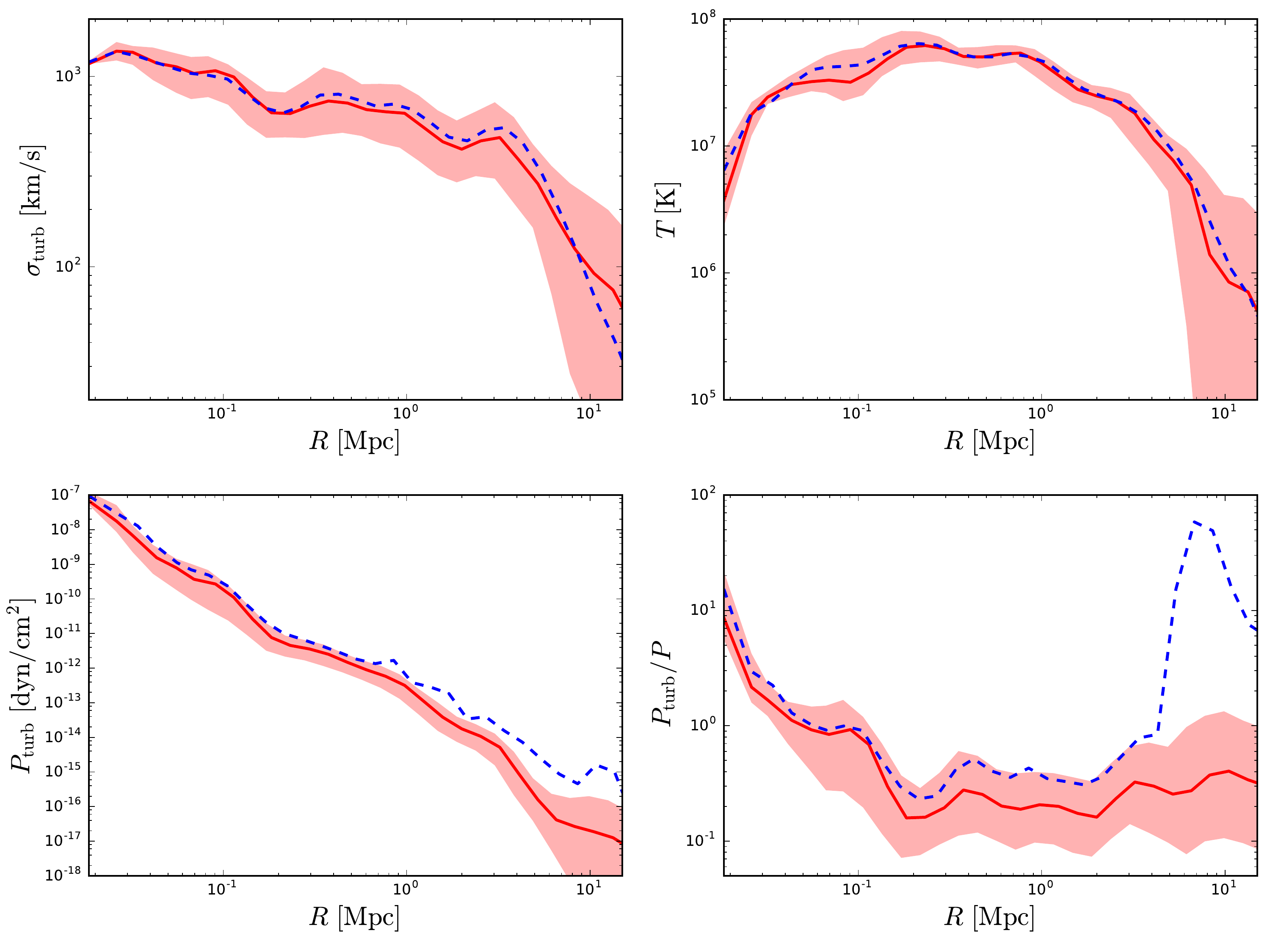}
    \caption{Radial profiles (volume-weighted: blue dashed, median: red solid, interquartile: red shaded) of the turbulent velocity dispersion (top left), temperature (top right), turbulent pressure (bottom left) and the ratio of turbulent and thermal pressures (bottom right) at redshift $z=0$.}
    \label{fig:turb}
\end{figure*}

Toward high wave numbers, numerically resolved modes are limited by the wave number $k_{\Delta}=\pi/\Delta$ in LES with resolution scale $\Delta$. In contrast to ILES, which simply assume $P(k_{\Delta})=0$, we have the lower bound
\begin{equation}
	P_{\rm sgs} := \frac{2}{3}\rho K \sim P(k_{\Delta})
\end{equation}
for the turbulent pressure on the grid scale. As shown by \citet{SchmFeder11,IapSchm11,IapiViel13,SchmAlm14}, $P_{\rm sgs}$ can reach a non-negligible fraction of the thermal pressure, although $P_{\rm sgs}/P_{\rm turb}$ is generally small compared to unity \citep{SchmEngl16}.\footnote{\citet{IapSchm11,IapiViel13} denote the turbulent pressure associated with SGS turbulence by $P_{\rm t}$, although this quantity depends on the arbitrary scale $\Delta$ set by the numerical resolution of a simulation, whereas $P_{\rm turb}$ encompasses all scales and does not \emph{per se} depend on resolution.} 
The expression for $P_{\rm sgs}$ is an exact result following from the filter formalism, on which LES are based \citep[see][]{Schmidt15}. However, there is no direct correspondence to the cumulative spectrum at a given wave number.

\begin{figure*}
	\includegraphics[width=\linewidth]{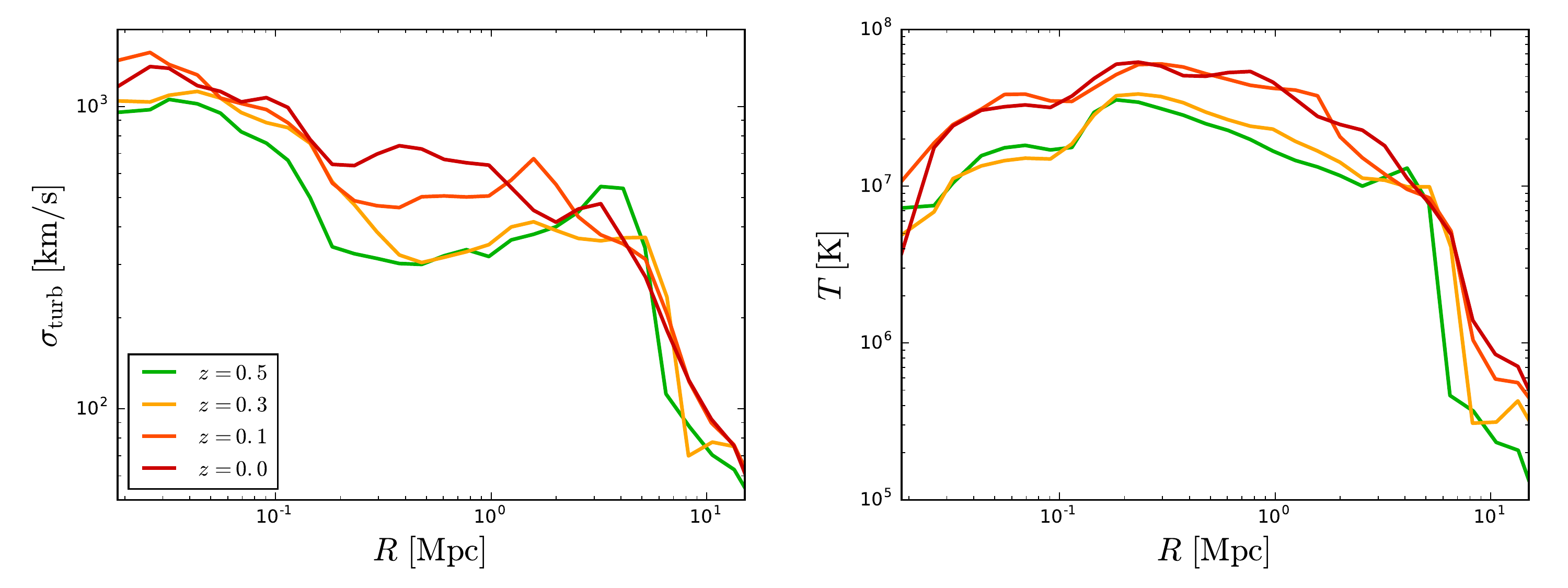}
    \caption{Medians of the turbulent velocity dispersion (left) and the gas temperature (right) at different redshifts.}
    \label{fig:turb_evol}
\end{figure*}

Profiles of the turbulent velocity dispersion and the turbulent pressure defined by equation~(\ref{eq:press_turb}) are shown in the left panels of Fig.~\ref{fig:turb}. While $\sigma_{\rm turb}$ shows the rather flat profile with a pronounced falloff at the outer shock fronts discussed in detail in \citet{SchmEngl16}, $P_{\rm turb}$ decreases gradually with radial distance from the centre of the cluster. This reflects mainly the decreasing gas density. As an indicator of the intensity of turbulence, $\sigma_{\rm turb}$ is the more useful quantity. Its evolution with redshift is shown in Fig.~\ref{fig:turb_evol}. The sudden increase of the turbulent viscosity in the aftermath of the major merger (see Fig.~\ref{fig:visc_evol}) is highlighted by $\sigma_{\rm turb}$, which grows from about $300\;\mathrm{km/s}$ at $z=0.5$ to roughly $600\;\mathrm{km/s}$ at $z=0$ in the WHIM. For the ICM, we see a lower relative increase. However, it is worthwhile to notice that the volume of the ICM is only a small fraction of the total turbulent gas volume in a cluster, assuming that the viscosity of the WHIM is sufficiently small.

The significance of turbulent pressure (defined one way or other) in clusters is commonly measured by the pressure ratio \citep[e.g.][]{VazzaBrun11,ParrCourt12b,IapiViel13,MiniBer15,VazzaWitt16}
\begin{equation}
	\frac{1}{\beta_{\rm turb}}:= \frac{P_{\rm turb}}{P} = \frac{\sigma_{\rm turb}^2}{3(\gamma-1)e},\,
\end{equation} 
where $e$ is the internal energy density and $\gamma$ the adiabatic coefficient. The dimensionless parameter $\beta_{\rm turb}$ is defined in analogy to the so-called plasma beta, which is given by the ratio of thermal and magnetic pressures. For our purpose, however, it is more suitable to work with the ratio of turbulent to thermal pressure. In the case $\gamma=5/3$ (which applies to our simulation), $P_{\rm turb}/P=\sigma_{\rm turb}^2/(2e)$ is identical to the ratio of the cumulative turbulent energy to the internal energy. Another parameter is the turbulent Mach number, which is defined by the ratio of the turbulent velocity dispersion to the speed of sound $c_{\rm s}$:
\begin{equation}
	\frac{\sigma_{\rm turb}}{c_{\rm s}} = \sqrt{\frac{3P_{\rm turb}}{\gamma P}}.
\end{equation}
Thus, we have $\sigma_{\rm turb}/c_{\rm s}=(1.8/\beta_{\rm turb})^{1/2}$ for monatomic gas. 

\begin{figure}
	\includegraphics[width=\linewidth]{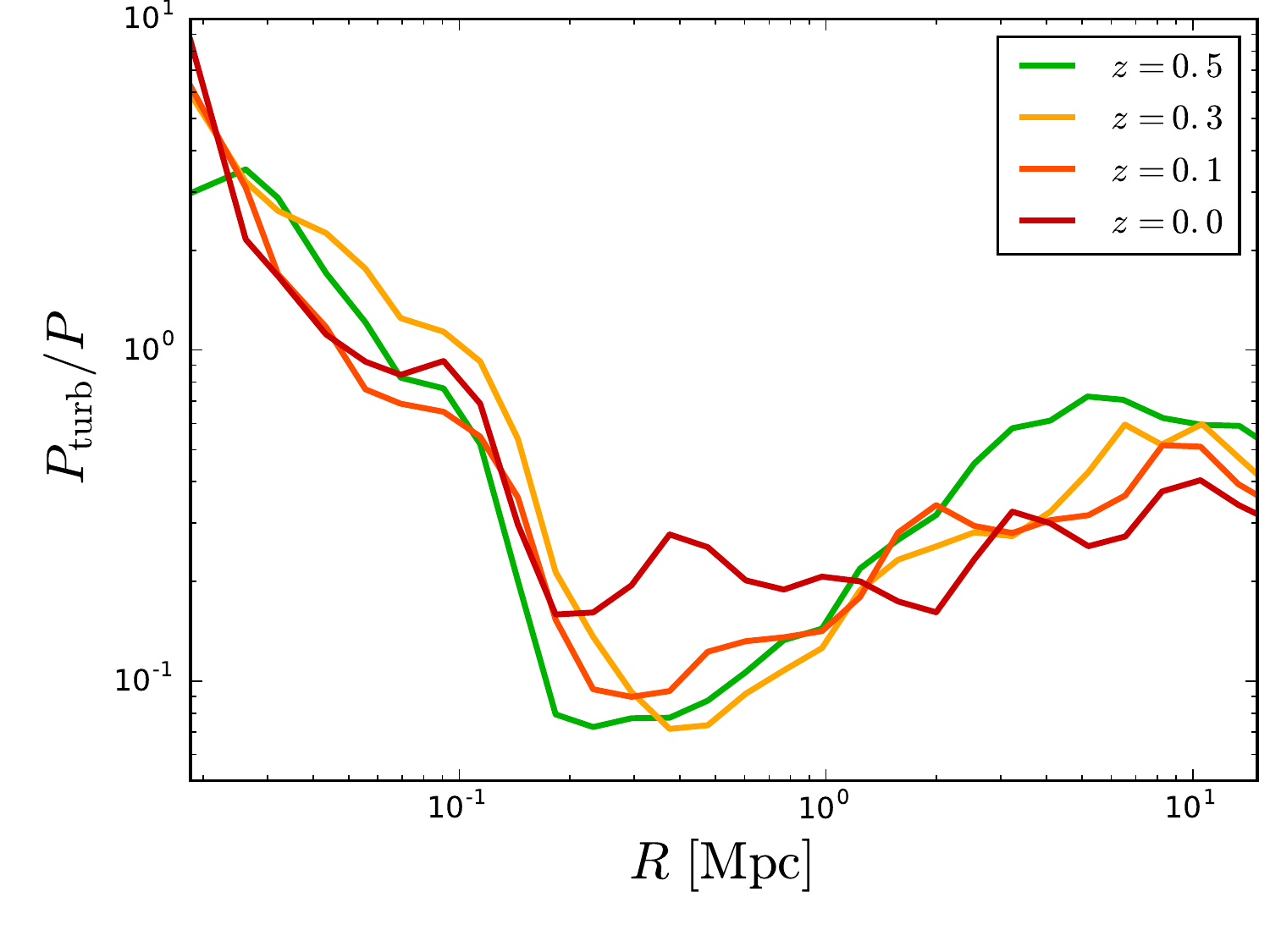}
    \caption{Evolution of the turbulent pressure relative to the thermal pressure. The graphs show radial profiles of the
    medians at different redshifts.}
    \label{fig:beta_evol}
\end{figure}

The volume-averaged and percentile profiles of $P_{\rm turb}/P$ at redshift $z=0$ are plotted in the bottom-right panel of Fig.~\ref{fig:turb}. As one can see, there is a large discrepancy between average and median values of $\sigma_{\rm turb}$, particularly close to the accretion shocks. The reason is revealed by considering the interquartiles of $\sigma_{\rm turb}$ and the temperature $T$ in the top panels of Fig.~\ref{fig:turb}. Especially at large radii, both quantities vary over several orders of magnitude. Since the turbulent pressure depends on $\sigma_{\rm turb}$, while the thermal pressure is given by $T$, very high values of $P_{\rm turb}/P$ can result in regions of low temperature and large turbulent velocity dispersion. These extreme values have a large impact on the average, but do not shift the median appreciably if they are rare. While the shell average of $P_{\rm turb}/P$ is about $0.4$ for $R\sim 1\;\mathrm{Mpc}$ and exceeds $10$ for radial distances around $10\;\mathrm{Mpc}$, we find a median between $0.2$ and $0.3$ for most of the WHIM, even in the vicinity of accretion shocks. Our results are roughly in agreement with previous studies by \citet{VazzaBrun11,ParrCourt12b,IapiViel13,VazzaWitt16}. According to our analysis, $P_{\rm turb}/P$ is typically higher in the cluster interior. This is not unexpected because we consider an exceptionally large and strongly turbulent cluster. Moreover, our method of computing the turbulent pressure is based on the three-dimensional velocity field, without assuming spherical symmetry or fixed integral scales of turbulence (see \citealt{SchmEngl16} for a detailed discussion of the different methods used in the literature). Here, we emphasise that the median values of $P_{\rm turb}/P$ increases much less in the cluster outskirts than the averages. As a consequence, the conclusion of \citet{ParrCourt12b} that turbulent pressure support can be a significant fraction of the thermal pressure for $R>R_{500}\approx 1.0\;\mathrm{Mpc}$ appears arguable if the distribution of $P_{\rm turb}/P$ is considered. However, the evolution of $P_{\rm turb}/P$ plotted in Fig.~\ref{fig:beta_evol} shows an increase of the median from about $0.1$ at $R\approx 0.2\;\mathrm{Mpc}$ to $0.7$ at several Mpc for $z=0.5$, which is closer to the profile shown in \citet{ParrCourt12b} (although they attribute turbulence production to the magnetothermal instability). Through the merger, $P_{\rm turb}/P$ becomes higher at radii below $1.0\;\mathrm{Mpc}$ and lower at larger radii, resulting in a much flatter profile at redshift zero. A qualitatively similar behaviour is reported in \citet{VazzaBrun11}, who compare merging to relaxed clusters.

\begin{figure*}
	\includegraphics[width=\linewidth]{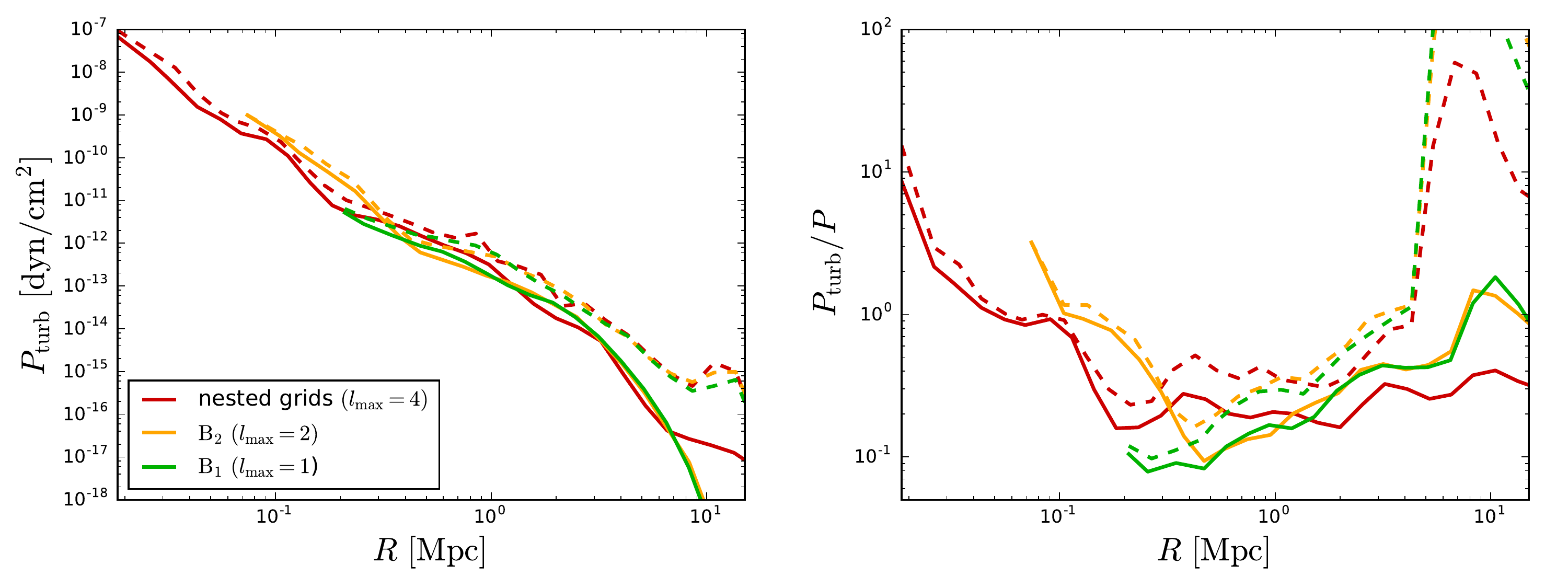}
    \caption{Comparison of radial profiles (volume-weighted: dashed, median: solid) of turbulent pressure
    (left) and the ratio of turbulent and thermal pressures (right) from different simulations. B$_1$ and B$_2$ are
    simulations with global AMR from \citet{SchmEngl16}. The root-grid resolution is $512^3$ in all simulations,
    the maximal number of refinement levels is indicated by $l_{\rm max}$ in the legend.}
    \label{fig:turb_res}
\end{figure*}

The strong increase of $P_{\rm turb}/P$ in the core, leading to a cusp-like central profile, is an indication of strong overcooling. Turbulent heating evidently does not counteract the overcooling. The central peak of $P_{\rm turb}/P$ becomes even more pronounced in the post-merger phase (see Fig.~\ref{fig:beta_evol}). The corresponding depression in the temperature profile at small radii (see Fig.~\ref{fig:turb_evol}) would likely be prevented by feedback from stars and AGNs at higher redshifts. Nevertheless, we will find it useful in the following section to consider the case of strong cooling without the restraining effect of feedback.

In Fig.~\ref{fig:turb_res} we compare profiles from our nested-grid simulation to simulations of lower resolution without nested grids. We chose the simulations B$_{1}$ and B$_{2}$ (see Table~1 in \citealt{SchmEngl16}), which have the same root-grid resolution as the nested-grid simulation. The spatial resolution at the highest refinement level is lower by a factor of $8$ and $4$ for B$_{1}$ and B$_{2}$, respectively. Moreover, the mass resolution is enhanced by a factor of $64$ in the nested-grid region, which encompasses the whole range of radii shown here. The left plot shows that the profiles of $P_{\rm turb}$ agree well in the overlapping radial ranges. For the ratio $P_{\rm turb}/P$, more pronounced variations with resolution can be seen. Particularly the core region is affected by the strong dependence of cooling on numerical resolution, which is also demonstrated by profiles of the specific internal energy shown in Fig.~15 in \citet{SchmEngl16}: the central depression in the internal energy profile becomes narrower and deeper with increasing resolution. As a result, the increase of $P_{\rm turb}/P$ toward the centre is shifted to smaller radii. As expected, the profile for B$_{2}$ is in between the profiles obtained for higher and lower resolutions. Outside of the core, however, there are systematic differences between the nested-grid simulation on the one hand and the two simulations with global refinement on the other hand, which cannot be fully explained by cooling. Apart from deviations in the median profiles of $P_{\rm turb}$, it appears that the thermal structure of the outer regions is altered in such a way that the radial dependence of $P_{\rm turb}/P$ is more levelled off in the nested-grid simulation. To a certain extent, this can result from changes in the structure of the cluster due to the improved resolution of lower-mass subhalos, which are particularly relevant in the outer parts. In addition, the drop of thermal and turbulent energies at the outer shock is less pronounced in the highest-resolution case. This is a consequence of the so-called flattening option in the hydrodynamical solver, which improves the stability at the cost of resolving strong jumps less sharply. This option was activated only in the nested-grid simulation (otherwise it would have been necessary to lower the CFL factor to unacceptably low values). We conclude that sensitivity to numerical resolution is clearly an issue, but the overall trends and typical values which are relevant for our discussion are sufficiently robust.

\section{Support}
\label{sc:support}

It is often argued that turbulent pressure contributes to the thermal support of the gas such that the effective pressure support is given by $P+P_{\rm turb}=(1+1/\beta_{\rm turb})P$ (an idea that dates back to \citealt{Chandra51}), which would imply an enhancement of $20$ to $30\;\%$ in the bulk of the WHIM. The notion that turbulence  generally stabilises self-gravitating objects against contraction can be found throughout the astrophysical literature. Compelling as it may seem, it was challenged by \citet{SchmColl13} in the context of supersonic turbulence in star-forming clouds by considering the local compression rate following form the substantial derivative of the divergence $d=\vecnab\cdot\vecU$. This approach was originally applied by \citet{ZhuFeng10} and further exploited by \citet{ZhuFeng11} and \citet{IapSchm11} for turbulence in the WHIM and ICM. The compression rate in the notation
introduced by \citet{SchmColl13} is given by\footnote{A derivation without SGS terms can be found in the appendix of \citet{ZhuFeng10}.}
\begin{equation}
  \label{eq:compr}
  -\frac{\DD d}{\DD t} = \frac{3H_0^2}{2a^2}\Omega_{\rm m}(\delta_{\rm m}-1) - \frac{1}{a}\Lambda + H d\,,
\end{equation}
where $a$ is the time-dependent scale factor, $H=\dot{a}/a$ the time-dependent Hubble parameter, $H_0$ the Hubble parameter for $z=0$, $\Omega_{\rm m}$ the constant (co-moving) density parameter of dark and baryonic matter, and $\delta_{\rm m}=\rho_{\rm m}/\rho_{\rm m, 0}$ the local overdensity of matter.
The local support function $\Lambda$ in co-moving coordinates $x_i$ encompasses terms related to thermal pressure,
\begin{equation}
  \label{eq:support_therm}
	\Lambda_{\rm therm} = -\vecnab\cdot\left(\frac{\vecnab P}{\rho}\right) =
	-\frac{1}{\rho}\frac{\partial^2 P}{\partial x_i\partial x_i} 
    + \frac{1}{\rho^{2}}\frac{\partial \rho}{\partial x_i}\frac{\partial P}{\partial x_i}\,,
\end{equation}
and numerically resolved turbulence
\begin{equation}
  \label{eq:support_turb}
	\Lambda_{\rm turb} =
	\frac{1}{2}\left(\omega^{2}-|S|^{2}\right)\,,
\end{equation}
where $\omega$ is the magnitude of the vorticity $\bmath{\omega} = \vecnab\times\vecU$ and $|S|^2=2S_{ij}S_{ij}$ the norm of the rate-of-strain tensor, which
is defined by the symmetrised velocity derivative: $S_{ij}=\frac{1}{2}(U_{i,j}+U_{j,i})$. Subgrid-scale dynamics gives rise to an additional term, which can
be split into the components
\begin{equation}
  \label{eq:support_sgs}
	\Lambda_{\rm sgs} =
	-\frac{1}{\rho}\frac{\partial^2 P_{\rm sgs}}{\partial x_i\partial x_i} 
    + \frac{1}{\rho^{2}}\frac{\partial \rho}{\partial x_i}\frac{\partial P_{\rm sgs}}{\partial x_i}
\end{equation}
and
\begin{equation}
  \label{eq:support_sgs_tf}
    \Lambda_{\rm sgs}^\ast =
    \frac{1}{\rho}\frac{\partial^2\tau_{ij}^\ast}{\partial x_i\partial x_j} 
	- \frac{1}{\rho^{2}}\frac{\partial \rho}{\partial x_i}\frac{\partial\tau_{ij}^\ast}{\partial x_j}
\end{equation}
corresponding to SGS turbulence pressure $P_{\rm sgs}=-\frac{1}{3}\tau_{ii}$ and the trace-free part $\tau_{ij}^\ast$ of the SGS turbulence stress tensor \citep[see][]{SchmEngl16}, respectively.
However, $\tau_{ij}^\ast$ is not accessible for postprocessing because only the numerical fluxes associated with the SGS terms in the dynamical equations are computed as part of the numerical integration scheme. This is why we have to resort to the trace $\tau_{ii}=-2\rho K=-3P_{\rm sgs}$ \citep[see][]{Schmidt15}, as proxy for SGS effects when analysing the support. Thus, we set
\begin{equation}
	\Lambda \simeq \Lambda_{\rm therm} + \Lambda_{\rm turb} + \Lambda_{\rm sgs}.
\end{equation}
for the total support in the following analysis. To calculate the derivatives in the support terms numerically, we applied centred finite differences of second order. For expressions such as $\vecnab P/\rho$ in eq.~(\ref{eq:support_therm}), face-centered density values (i.e.\ two-point averages) were used as weighing factors for the left and right pressure differences on the three-point stencil.

\begin{figure}
	\includegraphics[width=\linewidth]{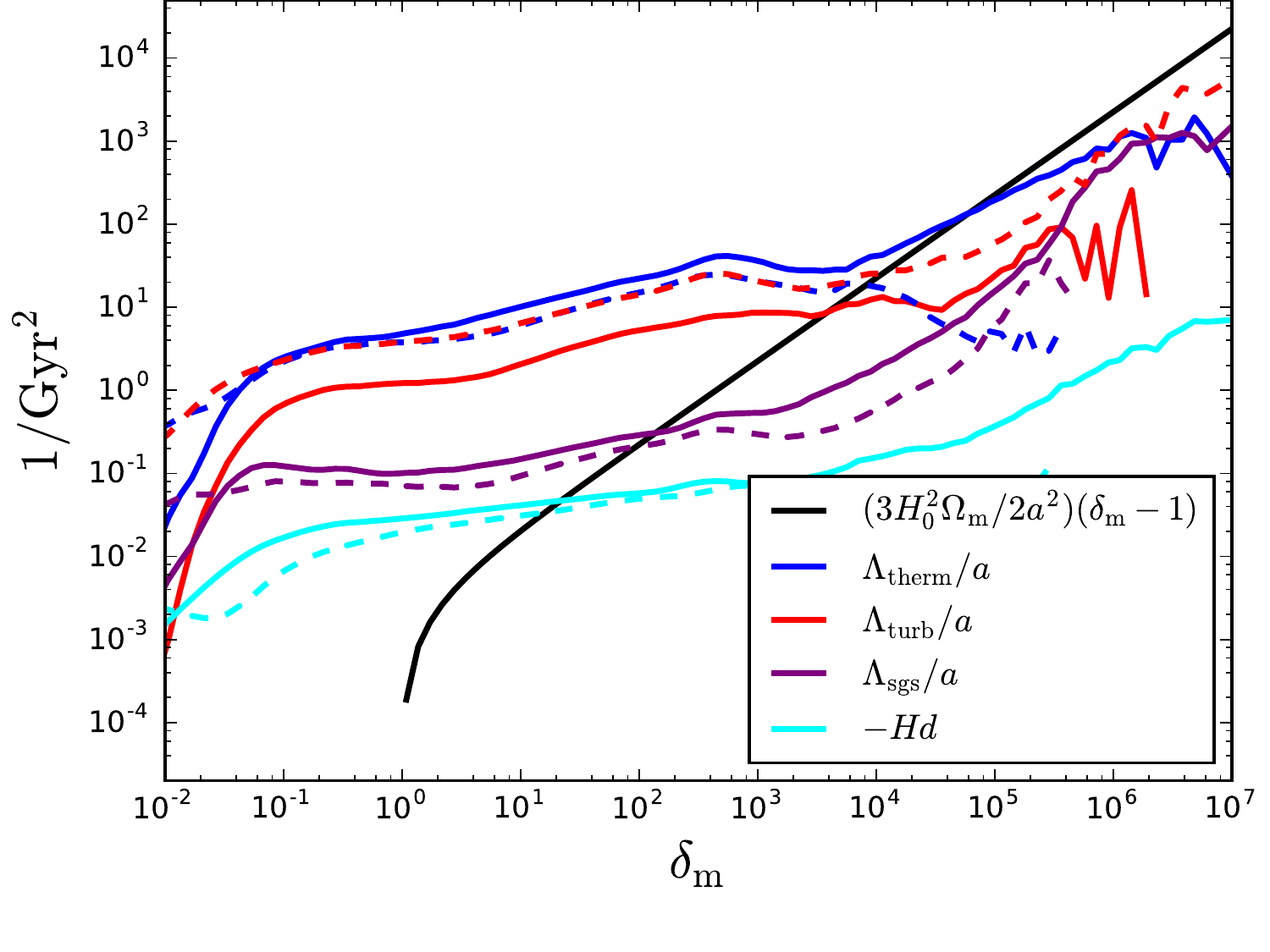}
    \caption{Mass-averaged profiles of different contributions to the compression rate vs.\ overdensity of dark and baryonic matter at $z=0$. Solid lines correspond to positive components, $\Lambda_+$, and dashed lines to the negative components, $\Lambda_-$. Different colours correspond to the terms indicated in the legend.}
    \label{fig:support}
\end{figure}

To compute statistics, we use the separation into positive and negative components introduced by \citet{SchmColl13}:
\begin{equation*}
	\Lambda_+ =\left\{\begin{array}{ll}
				\Lambda &\mbox{if}\ \Lambda \ge 0,\\
				0\ &\mbox{otherwise,}\ 
				\end{array}\right.\quad
	\Lambda_- =\left\{\begin{array}{ll}
				-\Lambda &\mbox{if}\ \Lambda \le 0,\\
				0\ &\mbox{otherwise.}\ 
				\end{array}\right.
\end{equation*}
Hence, $\Lambda_+$ is a source of divergence, while $\Lambda_-$ is a source of contraction (negative divergence). This separation has the additional merit of allowing us to plot statistics in logarithmic scaling. Figure~\ref{fig:support} shows both the positive (solid lines) and negative (solid lines) components of the support terms defined by equations~(\ref{eq:support_therm})--(\ref{eq:support_sgs}) for $z=0$. In addition, the gravity term and the cosmological expansion term in eq.~(\ref{eq:compr}) are plotted. Since all of these terms are source terms, we consider only averaged profiles. More specifically, mass averaging is applied. This is motivated by the fact that the divergence equation follows from the momentum equation by dividing through the density and applying the divergence operator. The average positive and negative components satisfy the relation $\langle\Lambda\rangle=\langle\Lambda_+\rangle-\langle\Lambda_-\rangle$. Therefore, the net support (total or any individual term) is positive if the solid line is above the dashed line, otherwise it is negative. As in \citet{ZhuFeng10,IapSchm11,SchmColl13}, bins of overdensity rather than radial distance are used. In this way, the profiles become much smoother and, expressing support vs.\ gravity sources, are physically more meaningful.

\begin{figure*}
	\includegraphics[width=\linewidth]{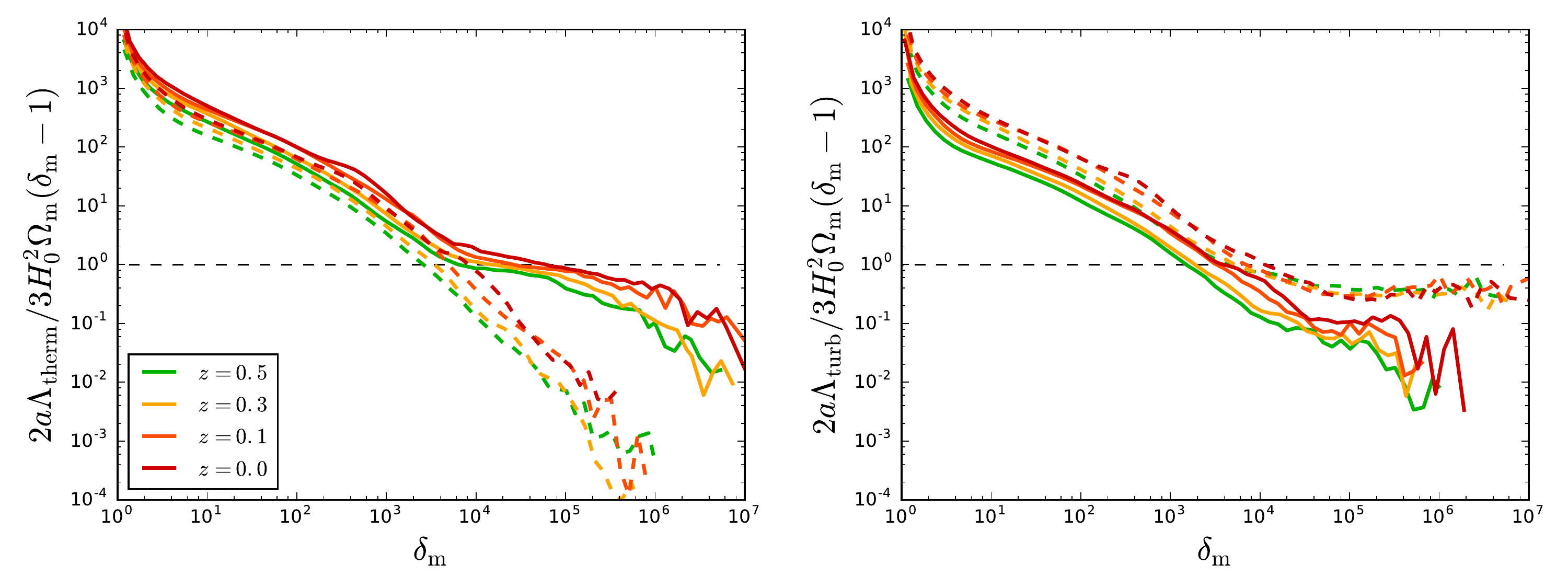}
    \caption{Mass-averaged thermal (left) and turbulent (right) support terms relative to the gravity term in eq.~(\ref{eq:compr}) for different redshifts. Solid lines correspond to positive components, $\Lambda_+$, and dashed lines to the negative components, $\Lambda_-$.}
    \label{fig:support_evol}
\end{figure*}

However, \citet{ZhuFeng10} and \citet{IapSchm11} consider only cells in which the support is positive (corresponding to $\Lambda_{\rm therm+}$ and $\Lambda_{\rm turb+}$ in our notation). From the distributions of the ratio $\Lambda_{\rm turb+}/\Lambda_{\rm therm+}$ versus overdensity, they find a mean ratio of order unity for the WHIM (i.e.~$\rho/\rho_0$ in the range from $\sim 10$ to several $100$). Apart from discarding all regions in which either $\Lambda_{\rm therm}$ or $\Lambda_{\rm turb}$ are negative, the mean values of $\Lambda_{\rm turb+}/\Lambda_{\rm therm+}$ tend to be biased toward high values (basically, for the reasons we discussed in the case of $P_{\rm turb}/P$ in Sect.~\ref{sc:press}). Indeed, profiles computed separately for $\Lambda_{\rm therm+}$ and $\Lambda_{\rm turb+}$ (solid blue and red lines in Fig.~\ref{fig:support}) show that the contribution from turbulence is lower. What is more, the profiles of the negative components $\Lambda_{\rm therm-}$ and $\Lambda_{\rm turb-}$ (dashed blue and red lines) reveal that they are about as important or even dominant. On the average, vorticity does not compensate for contraction due to the strain experienced by a fluid parcel, in agreement with analysis of forced supersonic turbulence in \citet{SchmColl13}. It is therefore meaningless to infer the impact of turbulence on the support of the gas solely on the basis of $\Lambda_{\rm turb+}$.\footnote{Vorticity and strain are given by the velocity derivative, which is dominated by velocity fluctuations on the smallest resolved scales. Consequently, it appears to be possible that turbulent eddies on larger scales yield stronger support. However, this was refuted by the spectral analysis for homogeneous supersonic turbulence in \citet{SchmColl13}, which shows that the net turbulent support is negative for all wavenumbers.} The SGS contribution is minor compared to $\Lambda_{\rm turb}$ (with the exception of the innermost core, which probably reflects an artefact of the model in this very small region at extremely high density). Even smaller is the contribution from the Hubble flow, which counteracts the contraction induced by gravity.\footnote{Neglecting the $\Lambda$-term, overdense gas ($\delta_{\rm m}>1$) produces negative divergence, $\updelta d<0$. The corresponding $H\,\updelta d$ is negative and reduces the compression rate due to gravity.} Since the Hubble time $1/H$ is of the order $10\;\mathrm{Gyr}$, $0.01\;\mathrm{Gyr}^{-2}$ is roughly the threshold above which the compression rate becomes dominated by local processes.

\begin{figure}
	\includegraphics[width=\linewidth]{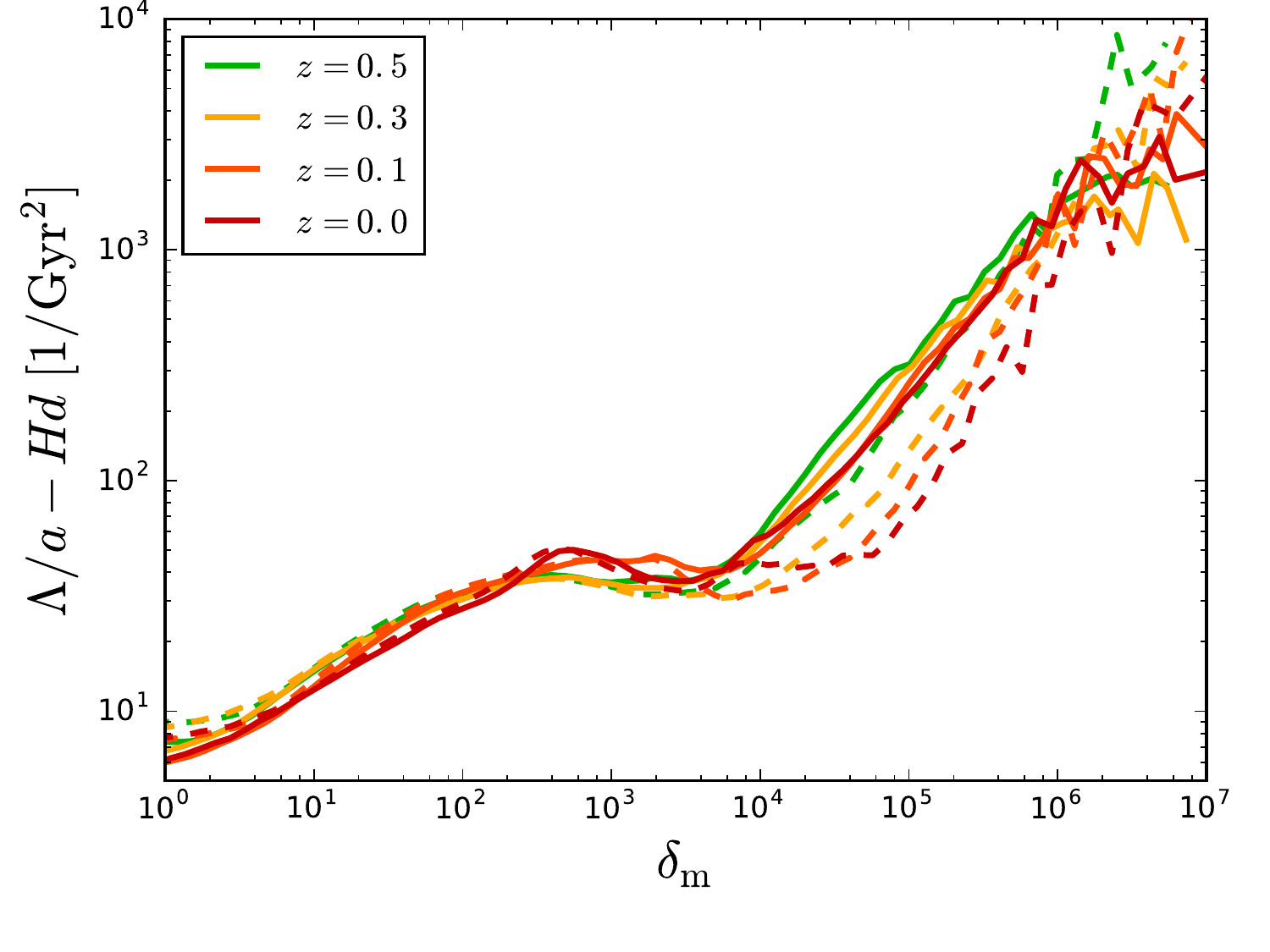}
 	\includegraphics[width=\linewidth]{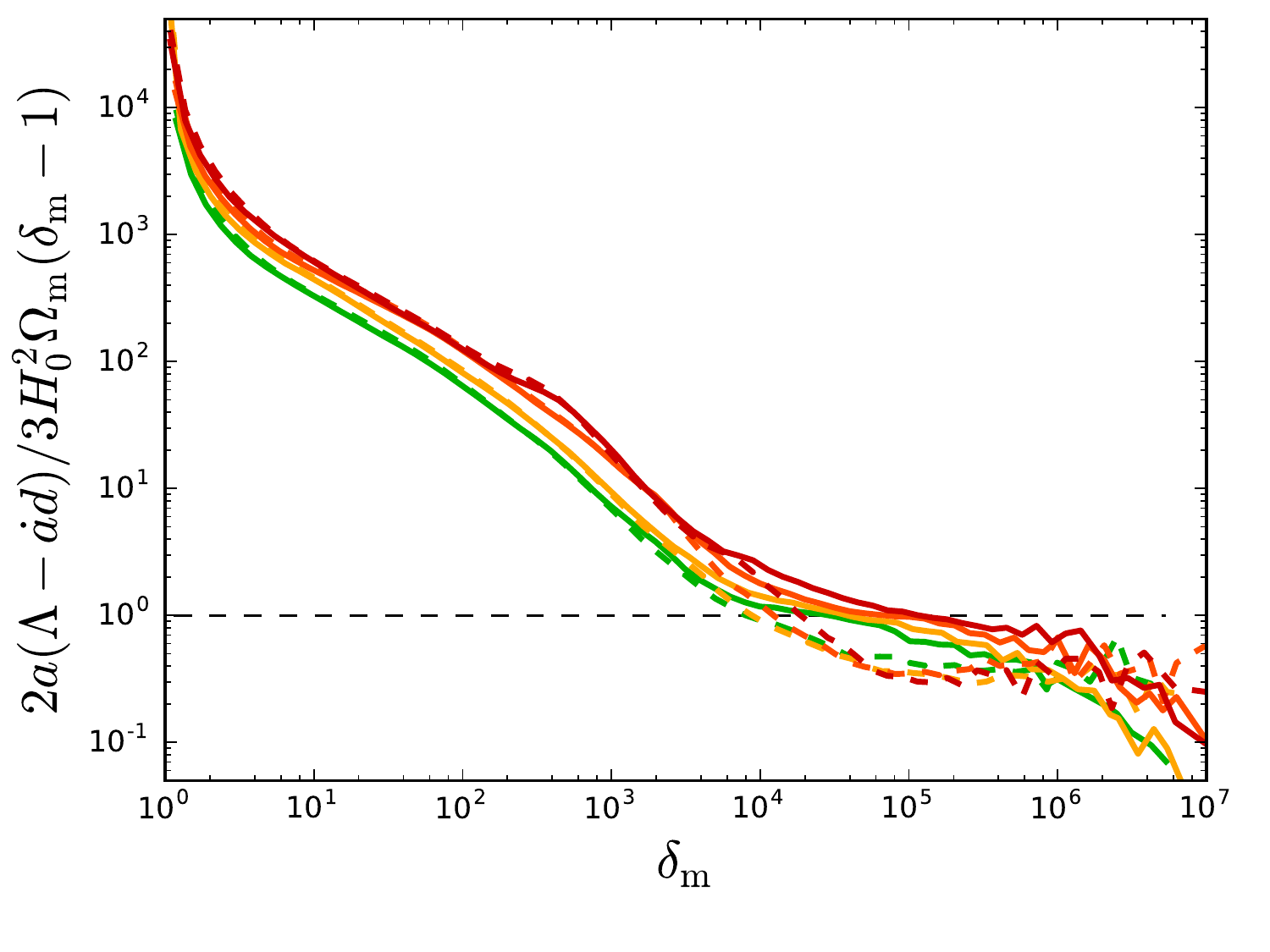}
   \caption{Mass-averaged net support $\Lambda=\Lambda_+-\Lambda_-$ plus cosmological expansion term for different redshifts. The bottom plot shows the net support relative to the gravity term. }
    \label{fig:support_tot}
\end{figure}

The evolution of the thermal and turbulent support terms relative to the gravity term is shown in Fig.~\ref{fig:support_evol}. Similar to the redshift-dependence of temperature and turbulent velocity dispersion (see Fig.~\ref{fig:turb_evol}), the thermal and the turbulent contributions to the support increase in the aftermath of the merger. In the case of $\Lambda_{\rm therm}$, the positive and negative components are more or less in balance for $\delta_{\rm m}\lesssim 10^3$, while $\Lambda_{\rm therm+}\gg \Lambda_{\rm therm-}$ for higher overdensities. Although $\delta_{\rm m}$ is the overdensity of both dark and baryonic matter, the lower and higher density ranges correspond quite well to the WHIM and ICM, respectively (this will be verified below). For the turbulent support, we find the opposite trend: on the average, $\Lambda_{\rm turb-}>\Lambda_{\rm turb+}$ at all densities, with a particularly strong dominance of $\Lambda_{\rm turb-}$ in the core. The two regimes become even more pronounced if the total support against gravity, i.e.\ $\Lambda/a-Hd$, is considered: in the top plot in Fig.~\ref{fig:support_tot}, one can see two distinct power-law ranges with a transition around $\delta_{\rm m}\sim 10^3$. 

\begin{figure*}
	\includegraphics[width=\linewidth]{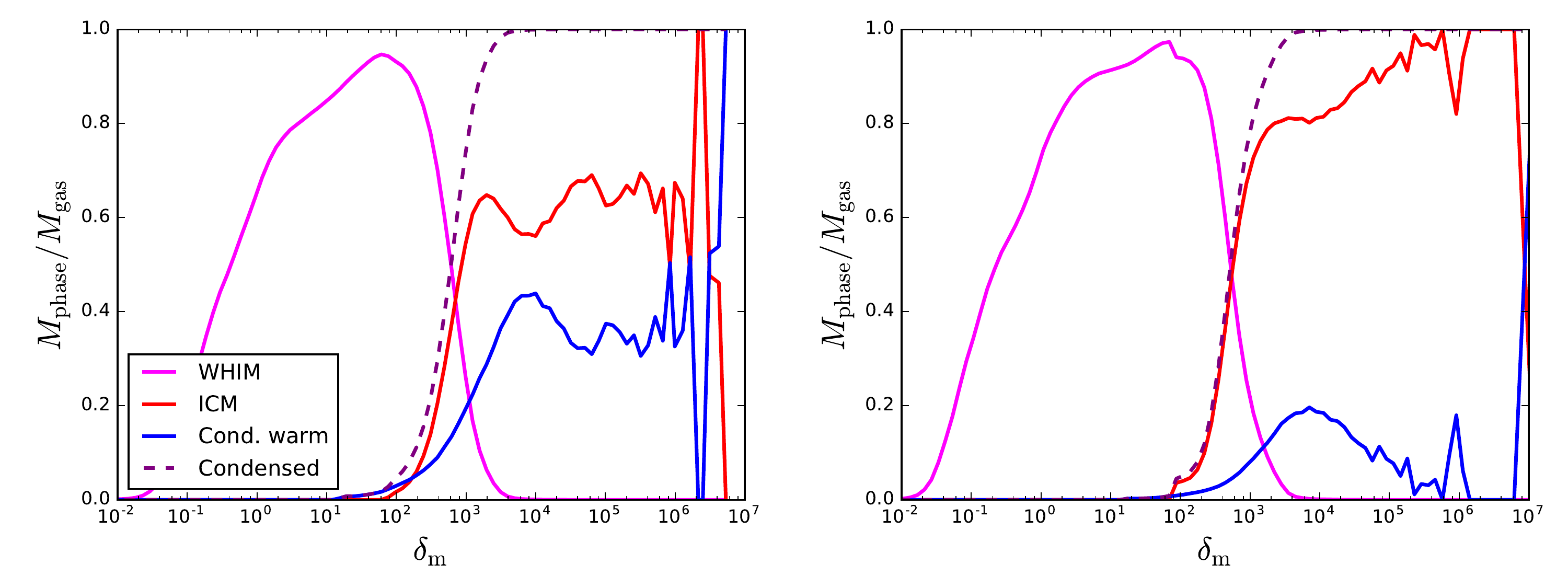}
    \caption{Mass fractions of different phases (WHIM: $\delta<500$ and $T>10^5\;\mathrm{K}$, ICM: $\delta>500$ and $T>10^5\;\mathrm{K}$, condensed warm: $\delta>500$ and $T<10^5\;\mathrm{K}$, condensed: $\delta>500$, where $\delta$ is the overdensity of baryonic gas; see \citealt{SchmEngl16}) for $z=0.5$ (left) and $z=0.0$ (right).}
    \label{fig:fract}
\end{figure*}

\begin{figure}
	\includegraphics[width=\linewidth]{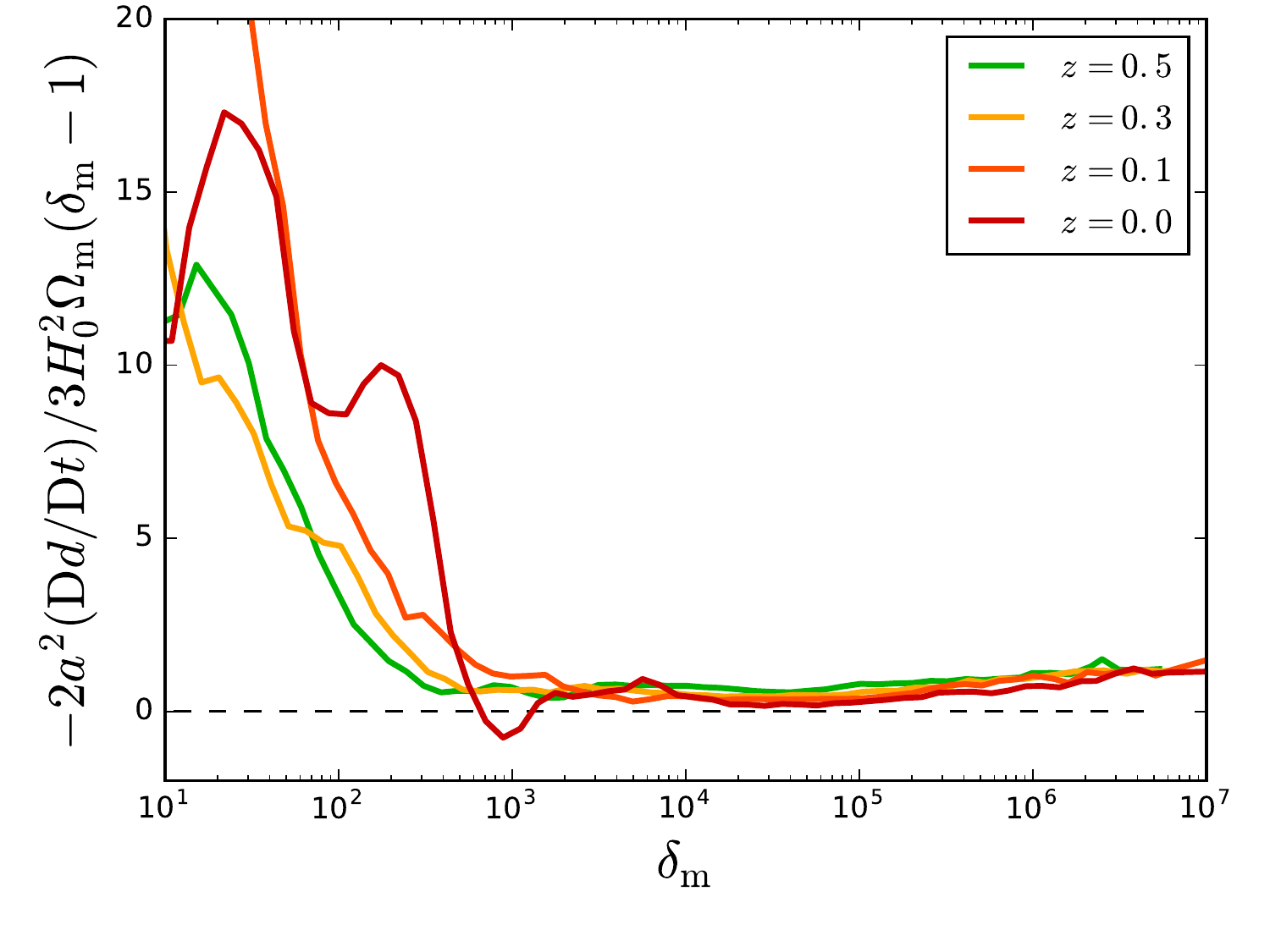}
    \caption{Total compression rate (see eq.~\ref{eq:compr}) normalized by the gravity term.}
    \label{fig:compr_rate}
\end{figure}

The magnitude of $2a^2\Lambda/3H_0^2\Omega_{\rm m}(\delta_{\rm m}-1)$ (see Fig.~\ref{fig:support_evol} and bottom plot in Fig.~\ref{fig:support_tot}) is indicative of the significance of self-gravity. In the low-density regime, which corresponds to the WHIM, the gas is only weakly self-gravitating. However, it is pulled toward the centre by the gravitational field of the dark-matter halo. This effect enters on the left-hand side of equation~(\ref{eq:compr}) through the peculiar velocity $\vecU$ in the advection term of the substantial time derivative:
\begin{equation}
	\frac{\DD d}{\DD t} = \frac{\partial d}{\partial t} + (\vecU\cdot\vecnab) d.
\end{equation}
In the high-density regime above $\delta_{\rm m}\sim 10^3$, on the other hand, the magnitude of the support is comparable to the gravity term, i.e.\ the gas is strongly self-gravitating. Assuming the definitions of \citet{SchmEngl16}, the mass fractions of the different phases plotted in Figure~\ref{fig:fract} demonstrate that the two regimes identified in Fig.~\ref{fig:support_tot} indeed correspond to the WHIM and ICM, respectively. Comparing redshifts $z=0.5$ and $0$, it can be seen that a substantial fraction of condensed gas with temperature $T<10^5\;\mathrm{K}$ (blue solid lines) is transferred into the ICM ($T>10^5\;\mathrm{K}$, red solid lines) by the merger, while the occupation of the WHIM (gas overdensity $\delta<500$)\footnote{The threshold overdensity $\delta=500$ for the ICM was motivated by the global phase diagram in \citet{SchmEngl16}. However, even if we had adopted the more common value of $10^3$, our basic conclusions would remain unaltered.} increases only by a small amount. The pronounced peaks at very high matter overdensities are caused by the condensation and cooling of gas in massive subhalos.

In Fig.~\ref{fig:compr_rate}, the compression rate normalised by the gravity term is plotted. The resulting profiles are mostly positive and below unity for $\delta_{\rm m}\gtrsim 10^3$, showing that the cluster core is close to equilibrium, albeit still contracting. The increase of the compression rate toward the centre is mainly caused by the cooling of the core. In the WHIM, we find a significantly higher compression rate after the merger ($z=0.1$ and $z=0$). This can be understood as a consequence of merger shocks. The multiple peaks that can be seen for $z=0$ indicate disturbances in the post-merger phase. 

\section{Conclusions}
\label{sc:concl}

We analysed thermal and turbulent properties of the gas in merging cool-core clusters in a cosmological simulation starting from nested-grid initial conditions. The baryonic gas was evolved with non-adiabatic physics,
including radiative cooling an a spatially homogeneous UV background, as detailed in \cite{LukStark15}. Refinement by overdensity and vorticity enabled us to resolve the ICM and the complete WHIM associated with the merging clusters in the nested-grid region down to $18\;\mathrm{kpc}$. A subgrid-scale model for numerically unresolved turbulence and an advanced filtering algorithm for the resolved flow \citep{SchmEngl16} were applied to calculate the turbulent velocity dispersion and the associated turbulent pressure. The main results are as follows.
\begin{itemize}
\item Particularly for the WHIM, we find that an assumed magnetic suppression factor $f\lesssim 10^{-3}$ for the effective viscosity \citep{GasChur13,RoedKraft13,SmithShea13,ZuHone15,SuKraft17} is necessary to allow for developed turbulence and to push the microscopic dissipation scale, which is approximately given by the Kolmogorov scale, sufficiently far below the grid resolution scale. The latter criterion is required for both the commonly applied ILES (implicit large eddy simulation with dissipation of purely numerical origin) and for the LES with explicit SGS model considered here. We estimate that an even stronger suppression of thermal conduction would be necessary to become negligible compared to turbulent diffusion (roughly by a factor of $10$ compared to the viscosity).
\item The major merger in our simulation both heats the gas and injects turbulence. Apart from some spatial redistribution, which is reflected by the flattening of radial profiles, the ratios of thermal and turbulent quantities are kept in balance in the course of the merger. This corresponds to the correlation between the mean thermal and turbulent energies of clusters found by \citet{MiniBer15,SchmEngl16}, which can be interpreted as a secondary manifestation of self-similar cluster evolution on top of the the self-similarity established for different halo masses.
\item It has been suggested that major mergers play a significant role in heating the cores of clusters in addition to feedback from AGNs. In agreement with previous studies \citep[e.g.][]{VazzaBruegg12,SkorHall13,LiBry15,HahnMar17}, however, we find that the merger by itself is insufficient to lift the cool cores of the predecessors into the regime of a non-cool core in the post-merger cluster. 
\item While the ratio $P_{\rm turb}/P$ of turbulent and thermal pressures suggests a substantial contribution of turbulence to the pressure support against gravity, the local compression rate of the gas reveals a different picture \citep{ZhuFeng10,IapSchm11}. Although the turbulent pressure becomes comparable to the thermal pressure in the cool core \citep{VazzaBruegg12,ParrCourt12b,IapiViel13,VazzaWitt16}, it turns out that thermal support is dominant. The turbulent support measured by the difference between the squares of vorticity and the rate of strain, is found to be effectively negative, similar to the results of \citet{SchmColl13} for supersonic turbulence in star-forming clouds. For the net support of the gas two regimes can be clearly identified, one for the WHIM and the other for the ICM. The latter is self-gravitating in the sense that the compression rate due to local density fluctuations is comparable in magnitude to the net support. The WHIM, on the other hand, is hydrodynamically dominated and characterised by an approximate balance between positive thermal and negative turbulent support. This can be understood as a consequence of the large strain induced by highly compressible turbulence, which overcompensates vorticity. In contrast to the indicators used by \cite{BifBor16}, which basically reflect the errors in shell-averaged accelerations for non-relaxed, merging clusters, we conclude that the support function $\Lambda$ in eq.~(\ref{eq:compr}) is similar in the pre- and post-merger stages.
\end{itemize}
Our findings regarding the viscosity indicate that first of all it is extremely important to achieve a better understanding of how the small-scale behaviour of a turbulent plasma is affected by magnetic fields \citep[see also][]{BrandSub05,EganShea16}. Implementing solvers for anisotropic diffusion terms in the equations of gas dynamics is maybe not the best route to follow because, from a microscopic point of view, there is no way of resolving the transversal suppression of transport coefficients in cosmological simulations. The notion of an effective diffusivity absorbs certain effects of turbulence and magnetic fields into the viscosity and conductivity of the plasma \citep{ChanCow98,NaraMed01,RuszOh10,ParrCourt12}. Consequently, microphysics and turbulent dynamics appear to be inextricably interwoven. For example, \citet{ZuHone15} showed that observed features such as cold fronts can be reproduced in simulations with a Braginskii-type anisotropic diffusion, which suppresses Kelvin-Helmholtz instabilities depending on the magnetic field strength. As a result, the fronts tend to be smoother than in simulations with strongly suppressed isotropic viscosity.

Another open question is whether the energetics remain self-similar if feedback is incorporated. This is not necessarily at odds with the apparent dichotomy of cool and non-cool cores, which refers only to the thermal state and entropy \citep{VazzaBruegg12,SkorHall13,GasBrig14,LiBry15,HahnMar17}. Since AGNs (or galactic winds in more sophisticated models) inject thermal energy and stir the gas in clusters \citep{Gaspari14}, it seems quite possible that the intensity of turbulence after episodes of AGN feedback is raised in such a way that the average $P_{\rm turb}/P$ remains unaltered (for example, an increase of vorticity due to AGN feedback is discussed in \citealt{LiBry14}). It is also conceivable, however, that variations in AGN activity bring about changes in $P_{\rm turb}/P$. 

While simulations help to gain a deeper understanding of the interplay between thermal and non-thermal physics in clusters, observational data for turbulence in clusters are still rare and, in most cases, rely on indirect detection methods. The situation is expected to improve with upcoming missions such as Athena. Compared to our simulation, the X-ray observations of the Perseus cluster by the \citet{Hitomi16} indicate a much lower intensity of turbulence in the innermost core (we find a turbulent velocity dispersion that is about an order of magnitude higher). Larger turbulent velocity dispersions in the subsonic regime were inferred by \citet{SandFab13,HofSand16}. However, there is clearly a tension between these data and the strong, transonic turbulence in our simulation. This could simply be a consequence of selection effects (we consider an exceptionally energetic and massive cluster), but it might also indicate the need for additional heating and/or damping mechanisms, which in the light of the theoretical uncertainties and model caveats discussed above do not seem altogether implausible.

\section*{Acknowledgements}

We thank the team of the Computational Cosmology Center at LBNL for developing and supporting \textsc{Nyx}.
J.~F.~Engels, C.~Behrens, and J.~C.~Niemeyer acknowledge financial support by the CRC 963 of the German Research Council. 
The simulation presented in this article was performed on the HLRN-III complex \emph{Gottfried} in Hannover (project nip00034). 
We also acknowledge the yt toolkit by \citet{TurkSmith11} that was used for our analysis of numerical data. 




\bibliographystyle{mnras}
\bibliography{cluster} 


\bsp	
\label{lastpage}
\end{document}